\documentclass{article}
\usepackage{amsfonts}
\usepackage{amssymb}
\usepackage{amsthm}
\usepackage{amsmath}

\usepackage{graphicx}

\usepackage{subfigure}
\usepackage{bm}             
\usepackage[mathscr]{eucal}
\usepackage{cancel}
\usepackage{psfrag}
\usepackage{wasysym}

\usepackage{oubraces}

\def\pmb#1{\setbox0=\hbox{$#1$}%
  \kern-.025em\copy0\kern-\wd0
  \kern.05em\copy0\kern-\wd0
  \kern-.025em\raise.0433em\box0}
\def\pmbs#1{\setbox0=\hbox{$\scriptstyle #1$}%
  \kern-.0175em\copy0\kern-\wd0
  \kern.035em\copy0\kern-\wd0
  \kern-.0175em\raise.0303em\box0}

\def\be{\begin{equation}}
\def\ee{\end{equation}}
\def\bea{\begin{eqnarray}}
\def\eea{\end{eqnarray}}

\def\Udot{\dot{U}}

\def\parb{\pmb{\partial}}

\def\hsp5{\hspace{5mm}}
\newcommand{\sfrac}[2]{\textstyle{\frac{#1}{#2}}}
\newcommand{\textfrac}[2]{{\textstyle{\frac{#1}{#2}}}}
\def\case#1/#2{\textstyle\frac{#1}{#2}}

\newcommand{\ue}{\check{u}}
\newcommand{\pe}{\tilde{p}}

\newcommand{\Thi}{\mathcal{T}_{\mathsf{Hi}}}
\newcommand{\Tlo}{\mathcal{S}_{\mathsf{Lo}}}
\newcommand{\Tco}{\mathcal{T}_{\mathsf{Jo}}}

\DeclareMathOperator{\sech}{sech}

\theoremstyle{plain}

\theoremstyle{remark}

\begin{document}

\title{\sc Spike Oscillations}
\author{
\sc J.\ Mark Heinzle$^{1}$\thanks{Electronic address: {\tt
mark.heinzle@univie.ac.at}}\ , 
Claes Uggla$^{2}$\thanks{Electronic
address: {\tt claes.uggla@kau.se}}\ ,\
and 
Woei Chet Lim$^{3}$\thanks{Electronic address: {\tt wclim@waikato.ac.nz}} \\
$^{1}${\small\em University of Vienna, Faculty of Physics,}\\
{\small\em Gravitational Physics, Boltzmanngasse 5, 1090 Vienna, Austria}\\
$^{2}${\small\em Department of Physics, University of Karlstad,}\\
{\small\em 65188 Karlstad, Sweden} \\
$^{3}${\small\em Department of Mathematics, University of Waikato,} \\
{\small\em Private Bag 3105, Hamilton, New Zealand \& }\\
{\small\em Albert-Einstein-Institut, Am M\"uhlenberg, 14476 Potsdam, Germany}}

\date{}
\maketitle
\begin{abstract}

According to Belinski\v{\i}, Khalatnikov and Lifshitz (BKL), a generic
spacelike singularity is characterized by asymptotic locality: 
Asymptotically, toward 
the singularity, 
each spatial point evolves independently from its neighbors, 
in an oscillatory manner 
that is represented by a sequence
of Bianchi type~I and~II vacuum models. Recent
investigations support a modified conjecture:
The formation of spatial structures (`spikes') 
breaks asymptotic locality. The complete
description of a generic spacelike singularity involves \emph{spike
oscillations}, which are described by sequences of Bianchi type I and certain
\emph{inhomogeneous\/} vacuum models. In this paper we describe how BKL and
spike oscillations arise from concatenations of exact solutions in
a Hubble-normalized state space setting, suggesting the
existence of hidden symmetries and showing that the results
of BKL are part of a greater picture.

\end{abstract}

\section{Introduction}

This paper is part of a research program on generic spacetime
singularities. It is concerned with \emph{spike oscillations}, a
missing piece in the description of the asymptotic dynamics
of spacetimes toward a generic spacelike singularity.

In research beginning in the 1960's, Belinski\v{\i}, Khalatnikov and
Lifshitz (BKL) claimed to have constructed an approximate general solution to
Einstein's field equations in the vicinity of a spacelike (`cosmological')
singularity~\cite{lk63,bkl70,bkl82}. The central assumption of BKL is that
the asymptotic dynamics is `\emph{local}' in the sense that
each spatial point evolves toward the singularity individually and independently of
its neighbors as a spatially homogeneous model. 
In addition, BKL gave heuristic arguments that suggested
that perfect fluid models with soft equations of state, such as dust or
radiation, are asymptotically `\emph{vacuum dominated}', i.e., asymptotically
toward the singularity the spacetime geometry is not influenced by the matter
content. Moreover, the BKL conjecture stated that for generic asymptotically vacuum
dominated models the evolution of each spatial point in the vicinity of the
singularity is approximated by a sequence of vacuum Bianchi type~I solutions 
(i.e., Kasner solutions) mediated by vacuum Bianchi type~II solutions, where the latter
determine a discrete `Kasner map' with chaotic
properties~\cite{khaetal85}--\cite{damlec11b}.
In general, the Kasner map generates infinite sequences of Kasner states,
and BKL argued that these sequences describe \emph{oscillatory}
temporal behavior toward a generic spacelike singularity.

In 1992 Belinski\v{\i} pointed out that 
the independent evolution of different timelines (due
to the locality assumption) leads 
to a lack of temporal synchronization of oscillations, and hence spatial
structure formation, reminiscent of turbulence, takes place, which has raised the
issue if the locality assumption of BKL is consistent~\cite{bel92}, see
also~\cite{monetal08} and references therein.

In 1993 Berger and Moncrief numerically studied vacuum $G_2$ models (i.e.,
models with two spacelike commuting Killing vectors~\cite{waiell97}) with
$T^3$ topology, so-called $T^3$ Gowdy models~\cite{gow71,gow74}, and observed
the development of large spatial derivatives near the singularity, which they
called `spiky features'~\cite{bermon93}. Furthermore, based on work by
Grubi{\v s}i\'c and Moncrief of the same year~\cite{grumon93}, these features
were tied to a  condition at isolated spatial surfaces that could
be imposed on a formal expansion of the Gowdy metric near the
singularity.\footnote{Due to the symmetries of $G_2$ models, fields only
depend on time and a single spatial coordinate; hence, two-dimensional symmetry 
surfaces, defined by a fixed value of this spatial coordinate, are often
referred to as spatial points in the $G_2$ literature. However, since our ultimate
objective is generic singularities (in spacetimes without symmetries), and 
because in that context this nomenclature is
inappropriate, we will refrain from referring to surfaces as
points.}

Toward the end of the 1990's, Berger, Moncrief and coworkers had found
further numerical evidence that the BKL picture seemed to be correct
generically in vacuum $G_2$ spacetimes, and even in spacetimes with only one
spacelike Killing vector, but that there were difficulties at exceptional
isolated spatial surfaces~\cite{bermon98}--\cite{beretal01}, see
also~\cite{ber02} and references therein. About the same time, Hern
independently resolved individual spatially spiky features to high numerical
accuracy, although for short time scales~\cite{her99,herste98}.

A significant analytic step toward an understanding of spiky features in
$T^3$ Gowdy models was accomplished in 2001 by Rendall and
Weaver~\cite{renwea01}. They applied a solution generating technique and 
Fuchsian methods
in~\cite{kicren98,ren00}, 
to produce
asymptotic expansions for `spikes', which they referred to as `true' or
`false spikes', where false spikes where shown to be gauge artefacts.
The work on spiky features in Gowdy spacetimes by Rendall and Weaver was
followed up by Garfinkle and Weaver in 2003 who used two different
complementary numerical techniques~\cite{garwea03}. In particular they
studied so-called `high velocity spikes' and found that they eventually
transform into `low velocity spikes'.\footnote{Velocities and other
diagnostic tools for describing and understanding asymptotic features
associated with singularities are discussed in Appendix~\ref{app:OT}. For
further and recent primarily analytic work on $T^3$ Gowdy models we refer to
the review~\cite{rin10} and references therein.}

Up to this point, the framework for studying singularities in
inhomogeneous spacetimes had been dominated by the metric approach, as
exemplified by the work of BKL, and the Hamiltonian approach used by Moncrief
and coworkers. 
However, at the same time, 
the studies of spatially homogeneous models were dominated by the
`Hubble-normalized' dynamical systems approach introduced by Wainwright and
Hsu~\cite{waihsu89}, which resulted in 
substantial progress~\cite{waiell97,rin00,rin01}.\footnote{For recent developments
about generic singularities in spatially homogeneous models,
see~\cite{heiugg09a}--\cite{lieetal10}.} In 2003 in an attempt to generalize this
framework to a general inhomogeneous context, Uggla {\em et
al\/}~\cite{uggetal03} (UEWE) recast Einstein's field equations into an
infinite dimensional dynamical system by means of `Hubble-normalized'
scale-invariant variables. This resulted in a more specific and precise formulation of the BKL conjecture 
in terms of the asymptotically regularized field equations
on an asymptotically bounded state space. 
The approach in UEWE was subsequently reformulated
somewhat by means of a conformal transformation~\cite{rohugg05} which led to 
a geometric framework that was in turn specialized by means of a conformally
Hubble-normalized Iwasawa frame by Heinzle {\it et al}~\cite{heietal09} to
provide a link to work on cosmological billiards~\cite{dametal03}, and to
establish the consistency of the BKL picture by means of dynamical systems
methods.

However, billiard and BKL consistency do not exclude other behavior at
generic singularities.
Numerical studies based on the dynamical systems formulation in UEWE gave
support for the BKL picture generically~\cite{andetal05,lim04}, and provided
evidence that the structure formation predicted by
Belinski\v{\i}~\cite{bel92} is not a threat to the consistency of the BKL
scenario, since it develops on superhorizon scale toward the singularity (see
also~\cite{gar04}). On the other hand, numerical and heuristic arguments
also suggested that there exist recurring temporal \emph{spike
oscillations} along certain timelines. These timelines correspond to spatial
points that form two-dimensional `\emph{spike surfaces}' and their
shrinking neighborhoods. The temporal spike oscillations are
gauge invariant features that are distinct from BKL oscillations, as
illustrated by the evolution of the Weyl scalars; in contrast to the BKL
case, spatial derivatives contribute significantly to the Weyl scalars along
the timelines of the spike surfaces (and in their shrinking neighborhoods), 
and in this sense oscillatory spike evolution is `non-local'.

Based on the solution generating technique used by Rendall and Weaver to
produce approximate solutions~\cite{renwea01}, in 2008 Lim was able to find a
1-parameter family of explicit (exact) inhomogeneous vacuum $G_2$
`\emph{spike solutions}', which are expressible in terms of elementary
functions~\cite{lim08}. As suggested by the work in~\cite{andetal05,lim04},
these solutions form the building blocks for spike oscillations. This was
further substantiated by Lim {\it et al} in 2009~\cite{limetal09} where the
analytic spike solutions found by Lim were numerically matched with high
accuracy to actual $G_2$ solutions in the regime toward the 
singularity (at least for a small number of spike oscillations).
Furthermore, Lim {\it et al} managed to provide numerical evidence that
showed that `higher order spike solutions,' with several spike surfaces,
could be described, locally and asymptotically, by the simpler `first order
spike solutions' with a single spike surface. Hence the first order spike
solutions seem to be the building blocks for describing spike oscillations in
general (at least for the spiky non-BKL behavior discovered so far in
numerical simulations), and it is therefore these solutions that we will
focus on in this paper.\footnote{For the influence of spike solutions on matter,
see~\cite{collim12}.}

In the BKL scenario, Bianchi type II vacuum solutions (`\textit{transitions}') connect Kasner
solutions, and thus generate the oscillatory BKL behavior through heteroclinic chains (BKL chains).
A specific goal of the present paper is to show how the explicit spike solutions by Lim
connect the (generalized) Kasner metrics in the Hubble-normalized state space
picture. Thereby we obtain a networks of `\emph{spike chains}' that represent
the asymptotic dynamics of solutions and thus the spike oscillations.
Throughout this paper we focus on general vacuum $G_2$ models.
There are reasons to believe that these
models describe essential features of generic spacelike singularities, i.e., the general singularities that
occur in models without symmetries~\cite{andetal05,limetal09}; we will 
return to this issue in the concluding remarks.

The paper is organized as follows. Section~\ref{sec:basics} provides a
succinct overview of the $G_2$ models in the conformal orthonormal (Iwasawa) frame
approach.
Section~\ref{sec:locnonloc} is devoted to the essential building blocks
of local BKL and non-local spike dynamics: Transitions. 
We discuss frame and curvature transitions and give a detailed
account of the explicit high and low velocity spike solutions
found by Lim tailored to our purposes and in the conformally 
Hubble-normalized picture.
Section~\ref{sec:OT} treats a special class of $G_2$ models,
the orthogonally transitive $G_2$ models which comprises, e.g., 
the $T^3$ Gowdy models. We describe in detail how
transitions can be \textit{concatenated} to obtain (finite) \textit{chains}
of transitions that are relevant for the asymptotic
dynamics of solutions.
Finally, section~\ref{sec:genG2} treats the general $G_2$ case. In
this case, concatenation leads to (in general) infinite BKL chains 
and spike chains that determine the oscillatory behavior of $G_2$ solutions
toward the singularity. 
Section~\ref{sec:concl} contains the conclusions 
together with a brief description and assessment
of previous numerical results in view of our present analytical and numerical
understanding. We also discuss connections between BKL and non-BKL behavior
and give hints to underlying hidden asymptotic symmetries. 
The appendices contain a discussion of on topological issues
and a brief review of the results on $T^3$ Gowdy spacetimes;
we stress that the `permanent' spiky features characteristic
of the $T^3$ Gowdy case are replaced by transient recurrent features
in the general $G_2$ case: Spike oscillations.

\section{$\bm{G_2}$ models in the conformal orthonormal frame approach}
\label{sec:basics}

In this paper we study vacuum $G_2$ models, which are characterized by the
existence of two spacelike commuting Killing vector fields. In
symmetry-adapted local coordinates these Killing vectors are $\partial_x$ and
$\partial_y$, which allows the metric to be written as
\begin{equation}\label{metricn}
d s^2 = -N^2 (dx^0)^2 + e^{-2b^1} \left(dx + n_1 d y  +
n_2d z\right)^2 + e^{-2b^2}
\left(dy  + n_3 d z\right)^2 + e^{-2b^3} dz^2\,,
\end{equation}
where the appearing functions depend on $x^0$ and the spatial coordinate $z$
alone, cf.~\cite{wai81}. With $n_1 = -\bar{n}_1$, $n_2 = -\bar{n}_2 +
\bar{n}_1 \bar{n}_3$, $n_3 = -\bar{n}_3$ we obtain the alternative
form~\cite{beretal01, beretal97, andetal04a, ren08}
\begin{equation}\label{metricbarn}
d s^2= -N^2 (dx^0)^2 + e^{-2b^1} \left(dx -\bar{n}_1 d y  +
(-\bar{n}_2 +\bar{n}_1 \bar{n}_3) \,d z\right)^2 + e^{-2b^2}
\left(dy - \bar{n}_3 d z\right)^2 + e^{-2b^3} dz^2\,.
\tag{\ref{metricn}${}^\prime$}
\end{equation}
We refer to Appendix~\ref{app:topology} for details and for a topological
discussion. The form~\eqref{metricn} provides an `\textit{Iwasawa}'
representation of the metric and is naturally associated with the orthonormal
spatial (co-)frame $\{\omega^1, \omega^2, \omega^3\}$ determined by
\begin{subequations}\label{iwa}
\begin{equation}
\begin{pmatrix} \omega^1 \\ \omega^2 \\ \omega^3 \end{pmatrix} =
\begin{pmatrix}
e^{-b^1} & 0 & 0\\
0 & e^{-b^2} & 0 \\
0& 0 & e^{-b^3}
\end{pmatrix}
\begin{pmatrix}
1 & n_1 & n_2 \\
0 & 1 & n_3 \\
0 & 0 & 1
\end{pmatrix}
\begin{pmatrix}
d x \\ d y \\ d z
\end{pmatrix},
\end{equation}
which is the dual of the orthonormal spatial `Iwasawa frame' $\{e_1, e_2, e_3\}$ (see, e.g.,
sections~2 and~13 in~\cite{heietal09}) given by
\begin{equation}
\begin{pmatrix} e_1 \\ e_2 \\ e_3 \end{pmatrix} =
\begin{pmatrix}
e^{b^1} & 0 & 0 \\
0 & e^{b^2} & 0 \\
0 & 0 & e^{b^3}
\end{pmatrix}
\begin{pmatrix}
1 & \bar{n}_1 & \bar{n}_2 \\
0 & 1 & \bar{n}_3 \\
0 & 0 & 1
\end{pmatrix}
\begin{pmatrix}
\partial_x \\ \partial_y \\ \partial_z
\end{pmatrix}.
\end{equation}
\end{subequations}
Combining this spatial frame with $e_0
= N^{-1} \partial_t$ and $\omega^0 = N d x^0$ yields an Iwasawa based
orthonormal frame for the spacetime metric, i.e., $d s^2 = -\omega^0 \otimes
\omega^0 + \delta_{\alpha \beta}\, \omega^\alpha \otimes \omega^\beta$, where
we use Greek letters as spatial frame indices.

The area density of the symmetry surfaces is $R = e^{-b^1-b^2}$. Whenever the
gradient of $R$ is timelike, it is possible to define time by
choosing $R$ to be a prescribed monotone function, e.g., $R \equiv x^0$, or
\begin{equation}\label{areatime}
-\log R =  b^1 +b^2 = t  + t_0\:,
\end{equation}
for some constant $t_0$ (which is chosen to be $t_0 = 0$ unless stated
otherwise). As shown in~\cite{beretal97, isewea03}, for the $T^3$ case,
$R\propto \exp(-t) \in (0,\infty)$ represents the maximal globally hyperbolic
development of initial data at $t = \mathrm{const}$, where $t\rightarrow
\infty$ represents the initial singularity; the only exception is the Taub
(flat Kasner) solution. In the \textit{area time gauge}~\eqref{areatime}, in
the nomenclature of~\cite{elsetal02}, the line element can be written as
\begin{equation}\label{metricarea}
d s^2 = e^{-2 b^3} ( -\bar{\alpha} e^{-2 t} d t^2 + d z^2)
+  e^{-2b^1} \big(dx + n_1 d y  + n_2 d z\big)^2 + e^{-2 t + 2 b^1}\!\left(dy + n_3 d z\right)^2 \:,
\end{equation}
where $\bar{\alpha} \neq 1$ in general.

An important subclass of the $G_2$ models are the \textit{orthogonally
transitive} (OT) models, see~\cite{car73,wai79} and p.~43 in~\cite{waiell97};
we refer to Appendix~\ref{app:topology} for details. In these models the
coordinates can be adapted so that $n_2 = n_3 = 0$ and $\bar{\alpha} = 1$,
which corresponds to setting the lapse function to%
\footnote{Throughout this paper we choose the direction of time toward the
past.}
\begin{equation}\label{lapse}
N =   -\exp(-[t + b^3]) = -\exp(-[b^1 + b^2 + b^3]) = -\sqrt{\det {}^{(3)}\!g}\,.
\end{equation}
The metric in the OT case thus simplifies considerably,
see also~\eqref{G2ot} and~\eqref{GowdyT3}.

Throughout this paper we focus on analyzing and representing solutions by
means of conformally Hubble-normalized Iwasawa adapted variables,
see~\cite{rohugg05} and section 2 of~\cite{heietal09}, rather than the metric
variables $b^1$, $b^2$, $b^3$ and $n_1$, $n_2$, $n_3$. The complete set of
relations between the conformally Hubble-normalized variables and the metric
variables are given in Appendix~\ref{app:methub}; here we restrict ourselves
to giving the relations for the key objects: The Hubble scalar, which is one
third of the derivative of the logarithm of the spatial volume density
w.r.t.\ proper time, is given by
\begin{equation}
H = -\textfrac{1}{3} N^{-1}\partial_{x^0}(b^1 + b^2 + b^3)\,,
\end{equation}
while the conformally Hubble-normalized shear variables take the form
\begin{subequations}
\begin{alignat}{2}
\Sigma_\alpha\, &\!\! := \Sigma_{\alpha\alpha} = -1 -
(HN)^{-1}\partial_{x^0}b^\alpha\qquad & &(\alpha = 1,2,3)\,, \\
\Sigma_{23} & =
\sfrac{1}{2}\exp(b^3-b^2)\,(HN)^{-1}\partial_{x^0}n_3\,, \qquad
& & \Sigma_{12} = \sfrac{1}{2}\exp(b^2 -
b^1)\,(HN)^{-1}\partial_{x^0}n_1\,,
\end{alignat}
\end{subequations}
so that $\Sigma_1 + \Sigma_2 + \Sigma_3 = 0$; note that $\Sigma_{31}=0$, see
Appendices~\ref{app:topology} and~\ref{app:methub}. We use Greek indices to
refer to the components of tensors w.r.t.\ the conformally Hubble-normalized
Iwasawa frame. In general a spatial orthonormal frame rotates w.r.t.\ a
gyroscopically fixed spatial frame (which is a so-called Fermi-propagated
frame). In the present approach, this spatial frame rotation is encoded in
the three functions $R_\alpha$, $\alpha =1,2,3$. As a consequence of the
Iwasawa parametrization, these functions are tied to the off-diagonal shear
components by
\begin{equation}
\Sigma_{23} = -R_1 \,, \qquad\Sigma_{12} = -R_3\,,\qquad \Sigma_{31} = R_2 = 0 \,,
\end{equation}
see~\cite{heietal09}. We choose to use $R_1$ and $R_3$ as
independent variables, instead of $\Sigma_{23}$ and
$\Sigma_{12}$. Note that $\Sigma_{23} = -R_1 = 0$ in the OT case, see Appendix~\ref{app:methub}.
Finally, it is of interest to also
give the Hubble-normalized spatial commutator function $N_1$ ($=N_{11}$):
\begin{equation}
N_1 = H^{-1}\exp(b^2 + b^3 - b^1)\partial_{z} n_1\,.
\end{equation}
The Einstein vacuum equation in the conformal Hubble-normalized orthonormal
frame approach are given in Appendix~\ref{app:fieldeqG2}.

In the following we will analyze $G_2$ vacuum models in terms of the
conformally Hubble-normalized orthonormal frame variables. In particular, to
illustrate the behavior of solutions, we will use projections onto
$(\Sigma_1, \Sigma_2, \Sigma_3)$-space and monitor the behavior of $N_1$
along the arising trajectories.

\section{Transitions}\label{sec:locnonloc}

In the Hubble-normalized approach a state space picture emerges which is
central for the analysis of the (asymptotic) dynamics of $G_2$ models. The
key structure on the boundary of the state space is the `Kasner circle' of
fixed points which represents `generalized' Kasner metrics. These fixed
points are connected by an intricate network of curves and families of
curves representing `generalized' spatially homogeneous solutions and
inhomogeneous solutions, which we denote by \textit{transitions} in this
context. The Kasner circle and the transitions turn out to be the building
blocks of generic oscillatory singularities of $G_2$ models
(and beyond); see section~\ref{sec:genG2}.

\subsection{The Kasner circle}

The Kasner, i.e. vacuum Bianchi type~I, solution is usually given as
%
\begin{equation}
ds^2 = - d\tilde{t}^2 + \tilde{t}^{\:2p_1} d\tilde{x}^2 +
\tilde{t}^{\:2p_2} d\tilde{y}^2 + \tilde{t}^{\:2p_3} d \tilde{z}^2\,,
\quad \big(p_1 + p_2 + p_3 = 1 = p_1^2 + p_2^2 + p_3^2\big)\,,
\end{equation}
where the constants $p_1, p_2, p_3$ are known as the Kasner
exponents. By making a constant transformation of the
coordinates according to
\begin{equation}
\tilde{t} = \hat{c}^0\!_0 \,\hat{t}\,,\qquad \tilde{x} = \tilde{c}^1\!_j \hat{x}^j\,,
\qquad \tilde{y} = \tilde{c}^2\!_j \hat{x}^j\,,\qquad \tilde{z} = \tilde{c}^3\!_j \hat{x}^j\,,
\end{equation}
the Kasner solution takes the form
\begin{equation}\label{genKasner}
ds^2 = - (\hat{c}^0\!_0 d \hat{t})^2 + \hat{t}^{\:2p_1}(\hat{c}^1\!_jd \hat{x}^j)^2 +
\hat{t}^{\:2p_2}(\hat{c}^2\!_jd \hat{x}^j)^2 + \hat{t}^{\:2p_3}(\hat{c}^3\!_jd \hat{x}^j)^2\,,
\end{equation}
where the constants $\hat{c}^i\!_j$ are obtained by appropriately scaling
$\tilde{c}^i\!_j$ with $\hat{c}^0\!_0$. 
The \emph{generalized} Kasner metric is defined by letting the constants
$p_\alpha$, $\hat{c}^0\!_0$ and $\hat{c}^i\!_j$ be arbitrary functions of the
spatial coordinates. As a consequence this metric is not a solution of
Einstein's field equations; its importance lies in its role as a building
block in the description of the asymptotic dynamics of actual solutions. In
the present context of $G_2$ models with Iwasawa parametrized metrics we have
$\hat{c}^2\!_1=\hat{c}^3\!_1=\hat{c}^3\!_2 =0$, and functions are functions
of $z$ ($=x^3$) alone.

In Appendix B of~\cite{heietal09} the conformal Hubble-normalized approach
was used to derive the generalized Kasner metrics by perturbing away from the
`\emph{local boundary}' (which previously was called the silent boundary,
see~\cite{limetal06, sanugg10} for a discussion). The local boundary is an
invariant boundary subset of the Hubble-normalized state space that is
obtained by setting all Hubble-normalized spatial frame components (and hence
spatial frame derivatives) to zero.\footnote{In the $G_2$ context the local
boundary is obtained by setting $E_3=0$, and hence $E_3\,\partial_z=
\parb_3 =0$, in Appendix~\ref{app:fieldeqG2}.} By construction, the equations
induced by the field equations on the local boundary coincide with the
equations describing spatially homogeneous models, where, however, constants
of integration become spatial functions. Accordingly, the local boundary
provides a well-defined state space setting for the BKL concept of
`generalized spatially homogeneous metrics'.

\begin{figure}[ht]
\centering
        \includegraphics[width=0.55\textwidth]{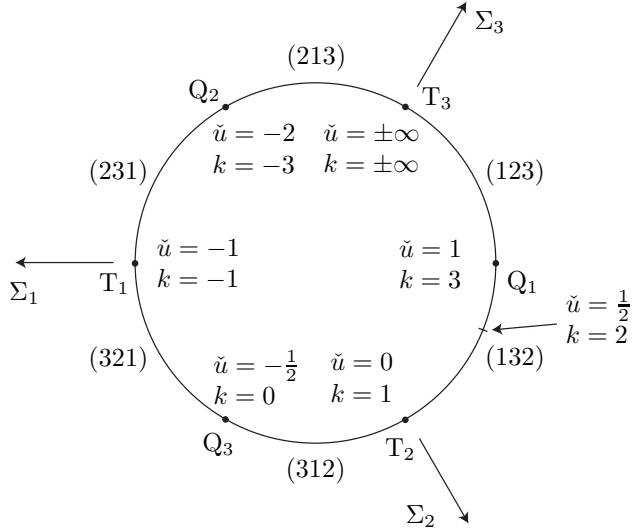}
\caption{The division of the Kasner circle $\mathrm{K}^{\ocircle}$ of fixed points
into six equivalent sectors and six LRS fixed points
$\mathrm{T}_\alpha$ and $\mathrm{Q}_\alpha$, $\alpha = 1,2,3$. Sector $(\alpha\beta\gamma)$ is defined by
$\Sigma_\alpha < \Sigma_\beta < \Sigma_\gamma$. The values of the extended Kasner parameter
$\ue$, cf.~\eqref{ueeq}, along $\mathrm{K}^{\ocircle}$ are indicated by the values of
$\check{u}$ at the LRS fixed points; in addition, $k = 2\ue + 1$ is given.}\label{fig:sectors}
\end{figure}

The structure on the local boundary that is of prime importance for our
understanding of the asymptotic dynamics of solutions toward a singularity is
the \textit{Kasner circle} $\mathrm{K}^\ocircle$. The Kasner circle is
obtained by setting all Hubble-normalized variables to zero except the
diagonal shear variables $\Sigma_1$, $\Sigma_2$, $\Sigma_3$. (In particular,
the rotation variables $R_1$, $R_2$, $R_3$ are zero, i.e., the frame that is used is a
Fermi-propagated frame.) The equations then imply that the diagonal shear
variables are constants in time but functions of the spatial coordinate(s).
The single remaining (algebraic) equation on the local boundary is the Gauss
constraint, which then characterizes $\mathrm{K}^\ocircle$:
\begin{equation}
\Sigma_\alpha = \mathrm{const}\,\,\forall \alpha,\qquad \Sigma_1 + \Sigma_2 + \Sigma_3=0\,,\qquad
\Sigma^2 := \sfrac{1}{6}\Sigma_{\alpha\beta}\Sigma^{\alpha\beta}  =
\sfrac{1}{6} \,(\Sigma_1^2 + \Sigma_2^2 +
\Sigma_3^2) = 1\,,
\end{equation}
where $\Sigma_\alpha$ are related to the generalized Kasner exponents
$p_\alpha$ according to\footnote{Constants are constants in time but
functions of the spatial coordinates; in the $G_2$ case of $z$. For this
reason the Kasner exponents are called `generalized' exponents.}
\begin{equation}
\Sigma_\alpha=3p_\alpha - 1\,\,\forall \alpha,\qquad p_1+p_2+p_3=1\,,
\qquad p_1^2 + p_2^2 + p_3^2 = 1\,.
\end{equation}
The Kasner circle is divided into six equivalent sectors, which are
associated with axes permutations and denoted by permutations of the triple
$(123)$: Sector $(\alpha\beta\gamma)$ is characterized by $p_\alpha < p_\beta
< p_\gamma$, see Fig.~\ref{fig:sectors}. The boundaries of the sectors are
six special points that are associated with locally rotationally symmetric
(LRS) solutions,
\begin{subequations}
\begin{alignat}{5}
\mathrm{T}_\alpha \,: \quad & (\Sigma_\alpha, \Sigma_\beta,
\Sigma_\gamma)\, &=&\, ({+2},{-1},{-1})\,, \quad &\text{or,
equivalently,} \quad &(p_\alpha,p_\beta,p_\gamma)\,
&=&\, (1,0,0)\,, \\
\mathrm{Q}_\alpha \,: \quad & (\Sigma_\alpha, \Sigma_\beta,
\Sigma_\gamma)\, &=&\, ({-2},{+1},{+1})\,, \quad &\text{or,
equivalently,} \quad &(p_\alpha,p_\beta,p_\gamma)\,
&=&\,(-\textfrac{1}{3},\textfrac{2}{3},\textfrac{2}{3})\,.
\end{alignat}
\end{subequations}
The \textit{Taub points} $\mathrm{T}_\alpha$, $\alpha = 1,2,3$, correspond to
the flat LRS solutions---the Taub representation of Minkowski spacetime,
while $\mathrm{Q}_\alpha$, $\alpha = 1,2,3$, yield three equivalent LRS
solutions with non-flat geometry.

A frame independent (gauge invariant) way of representing the Kasner
exponents $p_\alpha$ is by means of the (standard) Kasner parameter $u$; for
each of the six equivalent sectors $(\alpha\beta\gamma)$ we have
\begin{equation}\label{ueq}
p_\alpha = - u/f(u)\,,\qquad p_\beta = (1+u)/f(u)\,,\qquad
p_\gamma = u(1+u)/f(u)\,,
\end{equation}
where $u\in(1,\infty)$, while the boundary points of sector
$(\alpha\beta\gamma)$, $\mathrm{Q}_\alpha$ and $\mathrm{T}_\gamma$, are
characterized by $u=1$ and $u=\infty$, respectively. The function $f$, which
will be used extensively, see also e.g.~\cite{khaetal85}, is defined by
\begin{equation}
f = f(x) := 1 + x + x^2\,.
\end{equation}
In addition to the standard Kasner parameter $u$, we follow~\cite{damlec11a}
and define the \textit{extended} Kasner parameter $\ue$ by
\begin{equation}\label{ueeq}
p_1 = -\ue/f(\ue) \,,\qquad
p_2 = (1+\ue)/f(\ue) \,,\qquad
p_3 = \ue(1+\ue)/f(\ue)\,,
\end{equation}
where $\ue \in (-\infty,\infty)$ so that each value of $\ue$ distinguishes a
unique point on the Kasner circle, see Fig.~\ref{fig:sectors}.
Comparing~\eqref{ueq} and~\eqref{ueeq} we obtain a transformation between $u$
and $\ue$ for each sector:
\begin{subequations}\label{ueandu}
\begin{xalignat}{2}
& (123):\, (1,\infty) \ni \ue = u\,, \quad
& & (132):\, (0,1) \ni \ue = u^{-1}\,,  \\
&  (312):\, (-\textfrac{1}{2}, 0) \ni \ue =-\frac{1}{u+1}
\,,\quad
& & (321):\, (-1,-\textfrac{1}{2}) \ni \ue = -\frac{u}{u+1}\,, \\
&  (231):\, (-2,-1) \ni \ue =-\frac{u+1}{u}\,,\quad & &
(213):\, (-\infty,-2) \ni \ue = -(u+1)\,.
\end{xalignat}
\end{subequations}
%

%
%

%

%
%

\subsection{Frame transitions}

In the previous subsection the generalized Kasner metric is represented in a
Fermi propagated frame with diagonalized shear (which is at the same time an
Iwasawa frame). However, it is possible to represent the Kasner metric in a
rotating spatial frame (e.g., a rotating Iwasawa frame), which leads to a
time-dependent and non-diagonal Hubble-normalized shear. Since the frame
invariants of the shear tensor, i.e., its trace, square, and determinant,
must remain constants in time,  
the generalized Kasner metric in a (rotating) Iwasawa frame (for which
$R_2=0$) is characterized by the relations
\begin{subequations}\label{Kasnerrot}
\begin{align}
0 &= \Sigma_1 + \Sigma_2 + \Sigma_3\,,\label{Kasnerrota}\\
1 &= \Sigma^2 = \textfrac16 \left(\Sigma_1^2 + \Sigma_2^2 + \Sigma_3^2\big) +
\textfrac13 \big(R_1^2 + R_3^2\right) \,,\label{Kasnerrotb}\\
\mathrm{const} &= \det \Sigma_{\alpha\beta} =
\Sigma_1 \Sigma_2 \Sigma_3 - \Sigma_1 R_1^2 - \Sigma_3 R_3^2\,,\label{Kasnerrotc}
\end{align}
\end{subequations}
where the constant is merely a constant in time.

\begin{figure}[ht]
\centering
\includegraphics[height=0.4\textwidth]{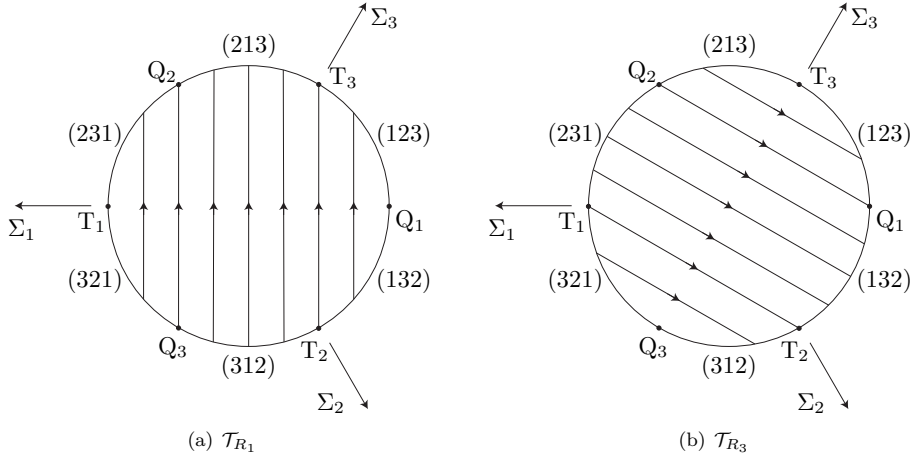}
        \caption{Projections of the two types of single frame transitions,
        $\mathcal{T}_{R_1}$ and $\mathcal{T}_{R_3}$,
          that exist in the $G_2$ case onto $(\Sigma_1, \Sigma_2, \Sigma_3)$-space.
          Throughout this paper, the direction of time, as indicated by the arrows,
          is toward the past.}
           \label{frametrans}
\end{figure}

A frame rotation in the 2-3-plane corresponds to $R_1 \neq 0$, $R_3 = 0$, and
thus, from~\eqref{Kasnerrotc}, by means of~\eqref{Kasnerrota}
and~\eqref{Kasnerrotb},
\begin{equation}
\Sigma_1 \left(\Sigma_2 \Sigma_3 - R_1^2\right)
= \Sigma_1 \left( \textfrac12 (\Sigma_2 + \Sigma_3)^2 + \textfrac12 \Sigma_1^2 - 3\right)
= \Sigma_1 \left( \Sigma_1^2 - 3\right)
= \mathrm{const}\,.
\end{equation}
Since $\Sigma_1 = \mathrm{const}$, this implies straight lines in $(\Sigma_1,
\Sigma_2, \Sigma_3)$-space, see Fig.~\ref{frametrans}(a). A single straight
line represents a Kasner solution (or a generalized Kasner metric 
corresponding to a constant value of $\check{u}$)
in a frame that is rotating in the 2-3-plane. A generalized Kasner metric 
associated with a spatially dependent $\check{u}$ corresponds to a family of lines.
Analogously, for a frame rotation in the 1-2-plane we have $R_1 = 0$, $R_3
\neq 0$, and the straight lines $\Sigma_3 = \mathrm{const}$, see
Fig.~\ref{frametrans}(b). We refer to the trajectories that individual frame
rotations give rise to as \emph{single frame transitions},
$\mathcal{T}_{R_1}$ and $\mathcal{T}_{R_3}$. Double frame transitions
$\mathcal{T}_{R_1 R_3}$, for which $R_1 R_3 \neq 0$, are not expected to play
an essential role in the asymptotic dynamics of solutions toward a generic
spacelike singularity, see~\cite{heietal09}, and we therefore refrain from a
discussion (the interested reader is referred to Appendix B and C
in~\cite{heietal09}). Note that transitions (i.e., solution trajectories) are
occasionally referred to as \emph{orbits}.

Linearizing the equations for $R_1$ and $R_3$ in Appendix~\ref{app:fieldeqG2}
at $\mathrm{K}^\ocircle$ gives
%
%
\begin{equation}\label{R13lin}
-\parb_0 R_1\big|_{\mathrm{K}^{\!\ocircle}}
= (\Sigma_2 - \Sigma_3)\!\big|_{\mathrm{K}^{\!\ocircle}} R_{1} =
3 (p_2 - p_3)R_1\, ,\qquad
-\parb_0 R_{3}\big|_{\mathrm{K}^{\!\ocircle}} =
3(p_1 - p_2)R_3\, .
\end{equation}
Hence the frame transitions $\mathcal{T}_{R_1}$ originate from sectors
$(321)$, $(312)$, $(132)$ and the points $\mathrm{Q}_3$ and $\mathrm{T}_2$,
where $R_1$ is unstable (w.r.t.\ a time directed toward the past), see
Fig.~\ref{frametrans}(a). We say that a frame transition $\mathcal{T}_{R_1}$
is `\textit{triggered}' by $R_1$; similarly, $R_3$ triggers the frame
transitions $\mathcal{T}_{R_3}$ in sectors $(213)$, $(231)$, $(321)$ and at
$\mathrm{Q}_2$, $\mathrm{T}_1$, see Fig.~\ref{frametrans}(b). 
Frame transitions map points of
$\mathrm{K}^\ocircle$ to each other; these maps can be expressed as the
following maps of the extended Kasner parameter $\ue$:
\begin{equation}
\mathcal{T}_{R_1}\!\!: \,\,\,
\ue \mapsto \ue^{-1},\quad\qquad \mathcal{T}_{R_3}\!\!: \,\,\,
\ue \mapsto -(\ue +1)\,.
\end{equation}
%

\subsection{Curvature transitions}

\begin{figure}[ht]
\centering{\includegraphics[height=0.5\textwidth]{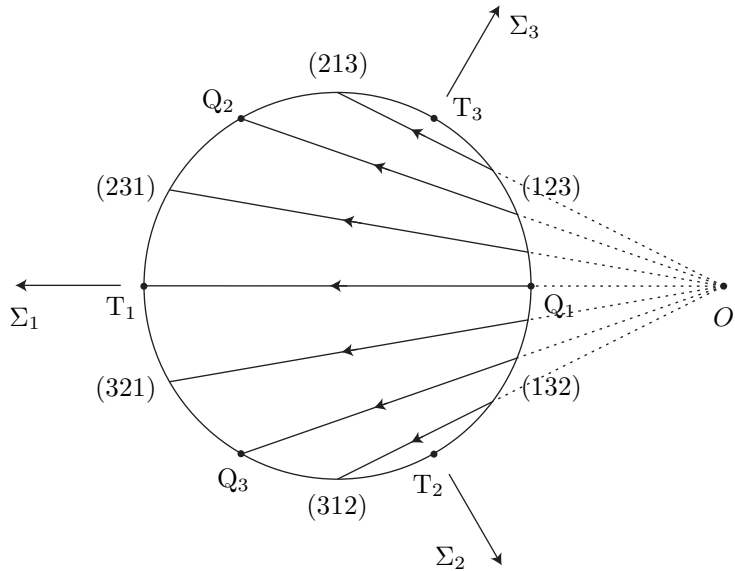}}
        \caption{Projections onto $(\Sigma_1, \Sigma_2, \Sigma_3)$-space
        of the single curvature transitions $\mathcal{T}_{N_1}$.}
    \label{curvtrans}
\end{figure}

The Bianchi type~II solutions on the local boundary are characterized by the
vanishing of all variables except for the Hubble-normalized shear variables
and the commutator function $N_1$. The simplest representation is w.r.t.\ a
Fermi propagated frame in which the shear has been diagonalized (which is an
Iwasawa frame at the same time), i.e., $R_\alpha = 0$ $\forall \alpha$. The
Bianchi type~II solutions can be conveniently parametrized as follows:
%
\begin{equation}
\label{typeIIlines}
\Sigma_1 = -4 + (1+ \ue^2 )\zeta\,,\qquad \Sigma_{\beta} =
2- \ue^2\zeta\,,\qquad \Sigma_{\gamma} = 2 - \zeta\,,
\end{equation}
where $\ue = \ue_- \in (0,+\infty)$ parametrizes the different possible
initial Kasner states 
and $\zeta$ is monotonically increasing,
see~\eqref{typeIIetaeq}. In the projection onto $(\Sigma_1, \Sigma_2,
\Sigma_3)$-space,~\eqref{typeIIlines} describes a family of straight lines
originating from the point \mbox{$(\Sigma_1,\Sigma_2,\Sigma_3)= (-4,2,2)$}
outside of $\mathrm{K}^\ocircle$, see Fig.~\ref{curvtrans}. At the same time,
the variable $N_1$ is determined by the constraint
\begin{equation}
\textfrac16 \left(\Sigma_1^2 + \Sigma_2^2 + \Sigma_3^2\right) + \textfrac{1}{12} N_1^2 = 1 \,.
\end{equation}
We refer to the type~II trajectories as \emph{single curvature transitions}
$\mathcal{T}_{N_1}$.

Linearizing the equation for $N_1$ in Appendix~\ref{app:fieldeqG2} at
$\mathrm{K}^\ocircle$ gives
\begin{equation}\label{N1lin}
-\parb_0 N_1\!\big|_{\mathrm{K}^{\!\ocircle}}
= -2( 1 + \Sigma_1)\!\left|_{\mathrm{K}^{\!\ocircle}}\right. N_1 = -6p_1 N_1 \,.
\end{equation}
%
Accordingly we say that the curvature transitions are `triggered' by $N_1$,
since $N_1$ is an unstable variable in sectors $(123)$ and $(132)$ and at the
point $\mathrm{Q}_1$ (w.r.t.\ a past-directed time variable), see
Fig.~\ref{curvtrans}.

The function $\zeta$ in~\eqref{typeIIlines} is governed by the equation 
%
%
\begin{equation}\label{typeIIetaeq}
\frac{\partial}{\partial( u t)}\, \zeta =
6\, |\ue_\pm|^{-1} \left(1 - \frac{\zeta}{\zeta_+}\right)\left(\frac{\zeta}{\zeta_-} - 1\right)\,,
\qquad \text{where}\quad \zeta_\pm
=\frac{3}{f(\ue_\pm)}\,,
\end{equation}
which has the solution
\begin{equation}
\label{zeta2} \zeta = \frac{3}{f(\ue_-)\frac{1-T}{2} +
f(\ue_+)\frac{1+T}{2}}\qquad\text{with}\quad T:= \tanh(2 u t)\,,
\end{equation}
so that $\zeta$ is monotonically increasing from $\zeta_-$ to $\zeta_+$.
By $u = u_-$ we denote the initial (standard) Kasner parameter; $\ue_\pm$ 
is the initial/final extended Kasner parameter.
It follows from~\eqref{typeIIlines} that a single curvature transition
$\mathcal{T}_{N_1}$ gives rise to a map
\begin{equation}
\mathcal{T}_{N_1}: \quad \ue = \ue_-  \mapsto \ue_+ = -\ue_- \:.
\end{equation}
In terms of the standard Kasner parameter $u$ this translates to $u_- \mapsto
u_+ = u_- -1$ if $u_- \geq 2$ and $u_- \mapsto u_+ = (u_- -1)^{-1}$ if $1\leq
u_-<2$, which is the usual representation of the Kasner map.

The Bianchi type~II metric assumes a less transparent form in an Iwasawa
frame that is rotating w.r.t.\ a Fermi propagated frame. In this case we
speak of mixed curvature/frame transitions since $N_1 \neq 0$ and $R_1$ or
$R_3 \neq 0$ simultaneously, see Appendix~D in~\cite{heietal09} for a
discussion. Although the trajectory of a mixed curvature/frame transition
looks rather complicated, the final state on $\mathrm{K}^\ocircle$ coincides
with the point obtained by successively applying the corresponding single
curvature transition and the relevant single frame transition(s); single
transitions thus act as building blocks that determine the final state of
multiple transitions. In any case, just like double frame transitions, the
mixed curvature/frame transitions are expected to not play a role in
asymptotic dynamics of solutions toward a generic spacelike singularity,
see~\cite{heietal09}.

\subsection{Spike transitions}

Together with $\mathrm{K}^\ocircle$ the frame and curvature transitions on
the local boundary constitute the essential building blocks for the
description of the asymptotic dynamics of $G_2$ models along `generic'
timelines, see sections~\ref{sec:OT} and~\ref{sec:genG2}. Evidently,
frame/curvature transitions and concatenations thereof concern the BKL
scenario of (asymptotic) locality in which the dynamics of timelines decouple
from their neighbors. However, to obtain a complete description of the
dynamics of $G_2$ models toward generic spacelike singularities, the failure
of asymptotic locality along particular timelines needs to be taken into
account. This failure of locality is associated with the formation and
recurrence of spikes; the central building block to describe this
\textit{non-local} scenario is the class of non-local transitions (`spike
transitions') associated with the `explicit spike solutions' found by
Lim~\cite{lim08}.

The \textit{explicit spike solutions} are $G_2$ OT
vacuum solutions (i.e., \mbox{$n_2=n_3=0$}) that in the area time
gauge~\mbox{\eqref{areatime}--\eqref{lapse}} take the form\footnote{The
solutions are presented here in a slightly different form than
in~\cite{lim08,limetal09}. Note also that, as stated in~\cite{limetal09},
there is a typographical error in equation~(34) in~\cite{lim08}. For further
discussion, see section~\ref{sec:OT}.}
\begin{subequations}\label{metricspike}
\begin{align}
b^1 &= \frac12\left[-t - \log\sech(w t) + \log\left(1 + \big(w e^t z\sech(w t)\big)^2\right) \right]\,,\\
b^2 & = t - b^1 \;, \\
b^3 &= \frac14\left[ \left( 3 + w^2 \right) t + 4\log\sech(w t) -
2\log\left( 1 +  \big(w e^t z\sech(w t)\big)^2\right) \right]\,,\\
n_1 &= \frac12 \,w\left[e^{-2t}+2\left( w \tanh(w t)-1\right) z^2\right]\,,
\end{align}
\end{subequations}
where $w > 0$.%
\footnote{It is possible to let $w \in\mathbb{R}$; however, $w \mapsto {-w}$
has the same effect as the coordinate reflection $y \mapsto -y$ and
thus results in the same solution, cf.~\eqref{coordtransf}; in
addition, the solution with $w = 0$ is trivial and not
associated with a transition since it corresponds to a Kasner
fixed point with $\ue = 1/2$.}

In the following we analyze the explicit spike solutions~\eqref{metricspike}
in terms of Hubble-normalized variables. For this purpose we define
\begin{equation}\label{tgpm}
T := \tanh (w t)\, ,\qquad \vartheta(T) := \left(1 -T\right)^{1-\frac{1}{w}}
\left(1 + T\right)^{1+ \frac{1}{w}}\, ,\qquad
g_\pm(T,z) := 1 \pm \vartheta(T) w^2 z^2\,.
\end{equation}
The time variable $T$ is bounded; $t\in(-\infty,\infty)$
corresponds to $T\in(-1,1)$. In the limit $T\rightarrow \pm 1$
we have
\begin{equation}
\lim_{T\rightarrow -1} \vartheta(T) = 0 \,,\qquad
\lim_{T\rightarrow +1} \vartheta(T) = \begin{cases} 0 & w >1 \,,\\ \infty & w < 1\,,\end{cases}
\end{equation}
and hence
\begin{equation}\label{gpmlims}
\lim_{T\rightarrow -1} g_\pm(T,z) = 1 \,,\qquad
\lim_{\substack{T\rightarrow +1 \\ w >1}} g_\pm(T,z) = 1 \,,\qquad
\lim_{\substack{T\rightarrow +1 \\ w <1}} g_\pm(T,z) = \begin{cases} 1 & z = 0 , \\
\pm \infty & z\neq 0 \,.\end{cases}
\end{equation}
Using the variable transformation given in
Appendix~\ref{app:methub} leads to
\begin{subequations}\label{pe}
\begin{align}
\label{pe1}
\pe_1 & := \frac{1}{3}\left(1 + \Sigma_1\right) =
\frac{2w\left(T - \frac{1}{w}\right)g_-(T,z)}
{(w^2+3)g_+(T,z) - 4w\left(T-\frac{1}{w}\right)}\,,\\[0.5ex]
\pe_2 & := \frac{1}{3}\left(1 + \Sigma_2\right) =
\frac{4g_+(T,z) - 2w\left(T - \frac{1}{w}\right)g_-(T,z)}
{(w^2+3)g_+(T,z) - 4w\left(T-\frac{1}{w}\right)} \,,\\[0.5ex]
\pe_3 & := \frac{1}{3}\left(1 + \Sigma_3\right) =
\frac{(w^2-1)g_+(T,z) - 4w\left(T - \frac{1}{w}\right)}
{(w^2+3)g_+(T,z) - 4w\left(T-\frac{1}{w}\right)}\,,
\end{align}
\end{subequations}
where  $\pe_1 + \pe_2 + \pe_3 = 1$, since $\Sigma_1 + \Sigma_2 + \Sigma_3 = 0$,
and where
\begin{equation}\label{N1R3spike}
N_1 = 12 w \frac{\sqrt{\vartheta(T)}\, z}{g_-(T,z)}\,\pe_1\,,\qquad
N_{12} = \frac{1}{2} \frac{\sqrt{1 - T^2}}{T - \frac{1}{w}}\, N_1\,,\qquad
R_3 = -3 \frac{\sqrt{1 - T^2}}{T - \frac{1}{w}}\,\pe_1\,.
\end{equation}
From these expressions it is straightforward to describe the
behavior of the explicit spike solutions~\eqref{metricspike}.
Henceforth, by `explicit spike solution'
we typically mean its representation in 
Hubble-normalized variables and its projection
onto $(\Sigma_1,\Sigma_2,\Sigma_3)$-space.

In contrast to frame and curvature
transitions, an explicit spike solution consists of an entire family of
curves parametrized by the spatial coordinate $z$.
As $t\rightarrow -\infty$ ($T\rightarrow {-1}$), each spike
solution~\eqref{metricspike} converges to a point on the Kasner circle
$\mathrm{K}^\ocircle$ described by the extended Kasner parameter $\ue = \ue_-
\in (0,+\infty)$, which we call the initial extended Kasner parameter, given
by
\begin{equation}\label{wue}
\ue = \ue_- = \frac12 \left(w + 1\right) \qquad \left( \Leftrightarrow \, w = 2 \ue - 1\right)\,.
\end{equation}
This relation becomes manifest when we use an alternative form
of~\eqref{pe} replacing $w$ by $2 \ue -1$, e.g.,
\begin{equation}\label{pe1ag}
\pe_1 =
\frac{- \left[{\ue}\left(\frac{1-T}{2}\right) + (1 - \ue)\left(\frac{1+T}{2}\right)\right]g_-(T,z)}
{f(\ue)\left(\frac{1-T}{2}\right) + f(1-\ue)\left(\frac{1+T}{2}\right)
+ f(\ue-1)\vartheta(T) w^2 z^2}\,.
\tag{\ref{pe1}${}^\prime$}
\end{equation}
Setting $T = -1$ and comparing with~\eqref{ueeq} establishes
$\ue$ as the initial extended Kasner parameter according
to~\eqref{wue}. Note that the requirement $w > 0$ corresponds
to $\ue > \textfrac{1}{2}$. If $\ue > 1$, it follows that the
initial sector is (123), while if $\textfrac{1}{2} < \ue < 1$,
the initial Kasner solution is in the part of sector (132) with
$1 < u < 2$ (where we recall from~\eqref{ueandu} that $\ue = u$
if $\ue >1$ and $\ue = u^{-1}$ if $0 < \ue < 1$).

\begin{figure}[ht]
\centering
        \includegraphics[width=0.85\textwidth]{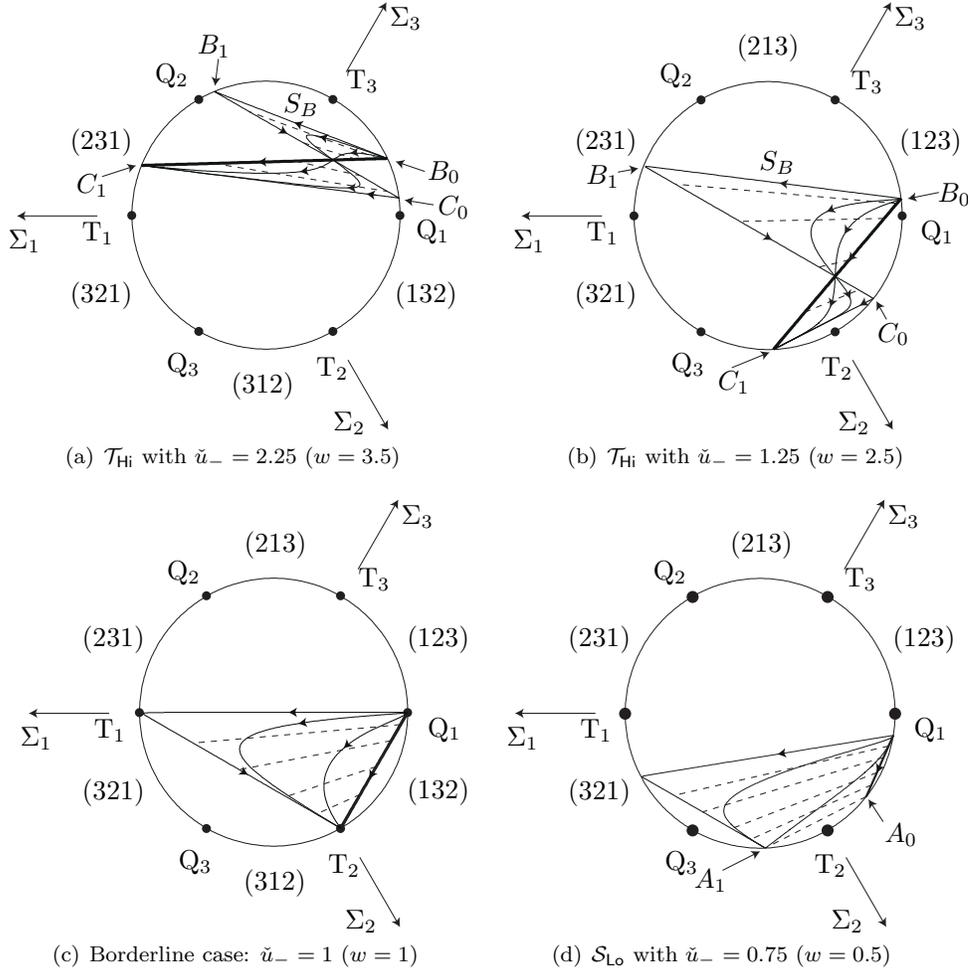}
        \caption{High velocity spike transitions $\Thi$ and low velocity spike solutions $\Tlo$ in their projections onto
          $(\Sigma_1, \Sigma_2, \Sigma_3)$-space. Each $\Thi$/$\Tlo$ corresponds to a family of curves
          parametrized by the spatial coordinate $z$.
          The thick (straight) lines represent the trajectories of the timelines of the
          spike surface $z = 0$, i.e., spike surface trajectories $\mathcal{T}_{S}$.
          The thin curves represent the trajectories of timelines $0 \neq |z| \not\gg 1$. The
          thin (straight) lines represent $|z| \gg 1$ timelines; these are short BKL chains,
          i.e., sequences of curvature/frame transitions.
          The dashed lines represent constant time slices of a solution.}
        \label{highandlowspiketranstime}
\end{figure}

\begin{figure}[ht]
\centering
        \includegraphics[height=0.37\textwidth]{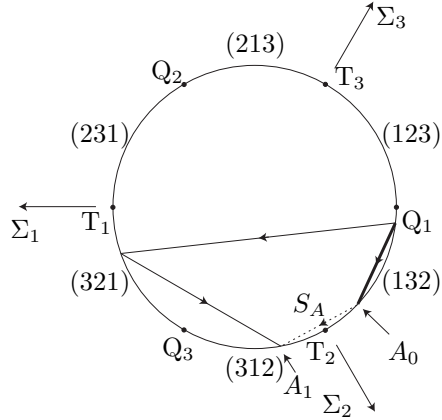}
        \caption{The `skeleton' of a $\Tlo$ spike solution described by
        the spike surface trajectory $\mathcal{T}_{S}$ (thick line) and the short
        BKL chain representing the trajectories associated with $|z| \gg 1$;
        cf.~Fig.~\ref{highandlowspiketranstime}; in addition the dashed line $S_A$ corresponds
        to a `missing' curvature transition $\mathcal{T}_{N_1}$ that would reunite the
        spike surface $z = 0$ with the $z \neq 0$ trajectories.}
        \label{Lowvel}
\end{figure}

In the limit $t\rightarrow \infty$ ($T\rightarrow 1$), however,
a qualitative difference arises that leads to a classification
of spike solutions in the OT $G_2$ context:
\begin{itemize}
\item[\textsf{Hi}] A spike solution with $w > 1$ (i.e.,
    $\ue = \ue_-
    > 1$) yields a \textit{`high velocity' spike transition}, which we
    denote by $\Thi$. (The term velocity is defined in
    Appendix~\ref{app:OT}.) Its limit as $t\rightarrow -\infty$
    ($T\rightarrow -1$) is a Kasner point in sector (123) with $\ue =
    \ue_- = (w+1)/2$, while its limit as $t\rightarrow \infty$
    ($T\rightarrow 1$) is a Kasner point that belongs to one of the four
    sectors $(2\beta\gamma)$, $(3\beta\gamma)$. This final Kasner point
    for the spike solution is characterized by an extended Kasner
    parameter $\ue_+$. Letting $T\rightarrow 1$ in~\eqref{pe1ag} shows
    that
\begin{equation}\label{spikemape}
\Thi: \quad \ue_+ = 1 - \ue_-\,.
\end{equation}
By means of~\eqref{ueandu}, the map~\eqref{spikemape} can be expressed in
terms of the standard frame-invariant Kasner parameter according to the
map:
\begin{equation}\label{spikemap}
\Thi: \quad u_+ =
\begin{cases}
u_- - 2 & u_- \in [3 ,\infty)\,, \\
(u_- -2)^{-1} & u_- \in [2,3] \,,\\
\big((u_- -1)^{-1} - 1 \big)^{-1} & u_- \in [ 3/2 ,2] \,,\\
(u_- -1)^{-1} - 1 & u_- \in [1, 3/2]\,,
\end{cases}
\tag{\ref{spikemape}${}^\prime$}
\end{equation}
where $u = u_-$ and $u_+$ are the initial and final Kasner parameters,
respectively. We refer to Figs.~\ref{highandlowspiketranstime}(a) and~\ref{highandlowspiketranstime}(b) for a
depiction of the projection of high velocity spike transitions onto
$(\Sigma_1,\Sigma_2,\Sigma_3)$-space.
\item[\textsf{Lo}] A spike solution with $0< w < 1$ (i.e.,
    $\textfrac{1}{2} < \ue  = \ue_- < 1$) yields a \textit{`low velocity'
    spike solution}, which we denote by $\Tlo$. Its limit as
    $t\rightarrow -\infty$ ($T\rightarrow -1$) is a Kasner point in the
    part of sector (132) with $\ue > \textfrac{1}{2}$ ($u <2$), which
    yields $u = u_- = 2/(w+1)$ since $u = \ue^{-1}$. Its limit as
    $t\rightarrow \infty$ ($T\rightarrow 1$) is a generalized Kasner
    metric with discontinuous Kasner parameter which is because the limit
    of $g_{\pm}(T,z)$ as $T\rightarrow 1$ is discontinuous,
    cf.~\eqref{gpmlims}. From~\eqref{pe1ag} it follows that
\begin{equation}\label{spikemaploe}
\Tlo: \quad \ue_+(z) =
\begin{cases}
1- \ue_-  & z = 0 \,, \\[0.5ex]
\ue_--1 & z \neq 0\,,
\end{cases}
\end{equation}
which in terms of the standard Kasner parameter results in
\begin{equation}\label{spikemaplo}
\Tlo: \quad u_+(z) =
\begin{cases}
u_- (u_- -1)^{-1} & z = 0 \,, \\[0.5ex]
(u_- - 1)^{-1} & z \neq 0\,.
\end{cases}
\tag{\ref{spikemaploe}${}^\prime$}
\end{equation}
Therefore the limit (as $t\rightarrow \infty$) of the low velocity spike
solution along the timelines of the spike surface $z = 0$ differs from the
limit along the other timelines. The trajectory in
$(\Sigma_1,\Sigma_2,\Sigma_3)$-space representing the spike surface ends in the
part of sector $(132)$ with $u > 2$ \mbox{($0 < \ue < \frac12$)}, while
trajectories associated with timelines with $z\neq 0$ meet in a point of
sector $(312)$; see Figs.~\ref{highandlowspiketranstime}(d) and~\ref{Lowvel}. This
behavior of the low velocity spike solution results in the convergence
of, e.g., the Hubble-normalized Kretschmann scalar to a discontinuous
limit.
\end{itemize}

The trajectories of an explicit spike solution that represent timelines with $|z| \gg
1$ (i.e., timelines far from the spike surface $z=0$) are approximated by a
succession of (two or three) curvature/frame transitions, see
Fig.~\ref{highandlowspiketranstime}. (In sections~\ref{sec:OT}
and~\ref{sec:genG2} we will refer to finite and infinite sequences of
curvature/frame transitions as \textit{BKL chains}.) In the case of a high
velocity spike transition $\Thi$, trajectories with $|z| \gg 1$ are
approximated by the chain $\mathcal{T}_{N_1} -\mathcal{T}_{R_3} -
\mathcal{T}_{N_1}$; in the case of a low velocity spike solution $\Tlo$,
$|z| \gg 1$ trajectories are approximated by a $\mathcal{T}_{N_1}$ followed
by a $\mathcal{T}_{R_3}$ transition. Formally, these chains are the limit as
$z \rightarrow \pm \infty$ of the orbits representing a spike solution. Since
$\mathcal{T}_{N_1}$ is associated with the map $\ue \mapsto -\ue$ while
$\mathcal{T}_{R_3}$ gives $\ue \mapsto -(\ue +1)$, it follows that
\begin{equation}
\textsf{Hi}: \quad
\ue_- \overset{\mathcal{T}_{N_1}}{\longmapsto} -\ue_-
\overset{\mathcal{T}_{R_3}}{\longmapsto} -(-\ue_- +1)
\overset{\mathcal{T}_{N_1}}{\longmapsto} 1 -\ue_- = \ue_+
\end{equation}
in the high velocity case, which coincides with~\eqref{spikemape}. In the low
velocity case the final $\mathcal{T}_{N_1}$ transition is missing so that
\begin{equation}
\textsf{Lo}:\quad
\ue_- \overset{\mathcal{T}_{N_1}}{\longmapsto} -\ue_-
\overset{\mathcal{T}_{R_3}}{\longmapsto} -(-\ue_- +1) = \ue_- -1 = \ue_+ \,,
\end{equation}
which coincides with the $z\neq 0$ case of~\eqref{spikemaploe}.

\begin{figure}[ht]
\centering{
        \includegraphics[height=0.37\textwidth]{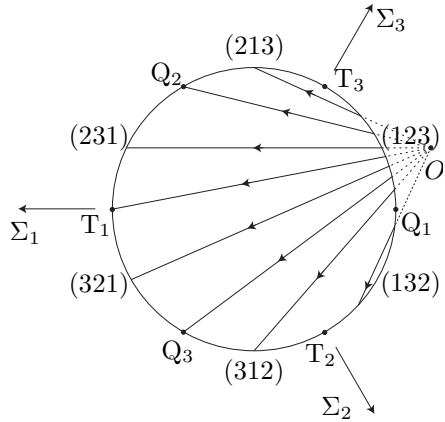}}
        \caption{Projections of spike surface trajectories
        $\mathcal{T}_{S}$ onto $(\Sigma_1, \Sigma_2, \Sigma_3)$-space.}
    \label{spiketrans}
\end{figure}

The evolution of the Hubble-normalized variables along the timelines of the
spike surface $z = 0$ itself is represented by a single straight line in
$(\Sigma_1,\Sigma_2,\Sigma_3)$-space, which we refer to as  
the `\emph{spike surface trajectory}' $\mathcal{T}_S$. 
The family of `spike surface trajectories'
(parametrized by $w$ or $\ue_-$) has a common focal point: Each spike surface
trajectory $\mathcal{T}_S$ corresponds to a straight line that originates
from the point $(\Sigma_1,\Sigma_2,\Sigma_3)= (-\frac52,\frac12,2)$ outside
$\mathrm{K}^\ocircle$, see Fig.~\ref{spiketrans}. In the high velocity case,
the spike surface trajectories enter the physical state space in sector
$(123)$, while in the low velocity case, they enter the physical state space
in the part of sector $(132)$ with $1< u < 2$, i.e., $\frac12 < \ue <1$.
From~\eqref{pe} we see that
\begin{subequations}
\begin{equation}\label{sigmaspike}
\Sigma_1\!\left|_{z=0}\right. = -\textfrac52 + \textfrac12 f(-\ue_\pm)\chi\,,\quad
\Sigma_2\!\left|_{z=0}\right. = \textfrac12 + \chi - \textfrac12 f(-\ue_\pm)\chi\,,\quad
\Sigma_3\!\left|_{z=0}\right. = 2 - \chi\,,
\end{equation}
where
\begin{equation}
\chi = \frac{3}{f(\ue_-)\frac{1-T}{2} + f(\ue_+)\frac{1+T}{2}}\,,
\end{equation}
\end{subequations}
so that $\chi$ increases monotonically from $\chi_- = 3/f(\ue_-)$ to $\chi_+
= 3/f(\ue_+)$. The expression $f(-\ue_\pm)$ appearing in~\eqref{sigmaspike}
stands for $f(-\ue_-)$ or $f(-\ue_+)$; either can be used since they
coincide.

Finally, the trajectories representing timelines $z \neq 0$ that are not far from the
spike surface `interpolate' between the two types of behavior, i.e., $|z| \gg
1$ and $z = 0$, described previously. This family of trajectories is rather
complicated, see Fig.~\ref{highandlowspiketranstime} and~\cite{lim08}; we
note, however, that all trajectories intersect at $T = 1/w$ in
$(\Sigma_1,\Sigma_2,\Sigma_3)$-space, as follows from~\eqref{pe}.


In equation~\eqref{metricspike} we have presented the principal form of the
explicit spike solution. There exist, however, entire families of explicit
spike solutions with different \textit{coordinate widths} and \textit{time
offsets}. These spike solutions are required when spikes are concatenated,
which we discuss in sections~\ref{sec:OT} and~\ref{sec:genG2}. We therefore
consider the constant coordinate transformation $(x, y, z) \mapsto (\bar{x},
\bar{y}, \bar{z})$ given by \mbox{$x = c^1\!_1\,\bar{x} + c^1\!_2\,\bar{y}$},
$y = c^2\!_2\,\bar{y}$, and $z = c^3\!_3\,\bar{z}$; in addition we may
consider a rescaling of time, i.e., $t \mapsto \bar{t}$ with $t =
c^0\!_0\,\bar{t}$. This coordinate transformation preserves the
form~\eqref{G2ot} of the metric provided that $c^0\!_0 = |
c^1\!_1c^2\!_2c^3\!_3|$ and induces the transformation $(b^1, b^2, b^3, n_1)
\mapsto (\bar{b}^1, \bar{b}^2, \bar{b}^3, \bar{n}_1)$ with
\begin{equation}\label{coordtransf}
\bar{b}^1 = b^1 - \log |c^1\!_1| \,,\quad
\bar{b}^2 = b^2 - \log |c^2\!_2| \,,\quad
\bar{b}^3 = b^3 - \log |c^3\!_3| \,,\quad
\bar{n}_1 = \frac{c^1\!_2}{c^1\!_1} + \frac{c^2\!_2}{c^1\!_1} \, n_1 \,.
\end{equation}
Choosing $c^1\!_1 = (2 Q_0)^{-1/2}$, $c^1\!_2 = (2 Q_0)^{-1/2} Q_2$, $c^2\!_2
= (2 Q_0)^{1/2}$, and $c^3\!_3 = \exp(-\lambda_2/4)$ reproduces the free
constants $Q_0$, $Q_2$, and $\lambda_2$ of Eqs.~(33)--(35) in~\cite{lim08}
and Eqs.~(A3)--(A5) in~\cite{limetal09}; note, however, that $\{c^1\!_1,
c^1\!_2, c^2\!_2, c^3\!_3\}$ are four free constants. Restricting ourselves
to the case $c^0\!_0 = 1$ (so that $| c^1\!_1c^2\!_2c^3\!_3| = 1$) it follows
from conceptual considerations or directly from the relations of this section
and Appendix~\ref{app:methub} that the Hubble scalar and the
Hubble-normalized variables are unaffected by~\eqref{coordtransf}, although
the frame variable $E_3$, cf.~\eqref{relations}, transforms like $E_3 \mapsto
\bar{E}_3 = (c^3\!_3)^{-1} \,E_3$. As a consequence, equations~\eqref{pe}
and~\eqref{N1R3spike} take the same form,
where $z \mapsto c_z z$, $c_z := c^3\!_3$, so that, e.g.,
\begin{equation}\label{gcz}
g_\pm(T,z) = 1 \pm \vartheta(T) w^2 c_z^2 z^2\,.
\end{equation}
%
%
Equation~\eqref{metricspike} with $z \mapsto c_z z$ yields a family of spike
solutions and transitions with different coordinate widths, an inverse
measure of which is, e.g., the constant $c_z$. It follows
from~\eqref{N1R3spike} that
\begin{subequations}\label{spikewidth}
\begin{equation}\label{spikewidth1}
c_z = \left(12 w \sqrt{\vartheta(T)}\,\pe_1|_{z=0}\right)^{-1}\, \partial_z N_1\,\big|_{z = 0} \,.
\end{equation}
To eliminate $T$ we make use of the trajectory of the spike surface $z = 0$,
because this trajectory is not affected by the spike width; using
$\pe_1|_{z=0}$, see~\eqref{pe1}, we obtain
\begin{equation}\label{spikewidth2}
\vartheta(T) 
= \frac{4}{w^2 (1 + 2 \pe_1|_{z=0})^2} \,\big( {-\ue_+} - f(\ue_+) \pe_1|_{z=0}\big)^{1-\frac{1}{w}}
\big(\ue_- + f(\ue_-) \pe_1|_{z=0}\big)^{1+\frac{1}{w}}\,,
\end{equation}
\end{subequations}
where $\ue_-$ ($\ue_+$) is the initial (final) point of 
$\Thi$ or $\Tlo$ (for $z = 0$), i.e., $\ue_+ = 1- \ue_-$, $w = 2 \ue_- -1$.
Inserting~\eqref{spikewidth2} into~\eqref{spikewidth1} we obtain an explicit
formula for $c_z$ in terms of certain Hubble-normalized variables (and a
spatial derivative).

While a rescaling of the spatial coordinate $z$ yields a spike solution with
a different coordinate width, a translation in $z$ puts the spike surface at
a different spatial location. Analogously, translations in time (`time
offsets'), i.e., $t\mapsto t \pm t_0$, yield spike solutions with a different
temporal flow (which, of course, does not affect the orbits in the diagrams).
In this case, the `time' $T$ of~\eqref{tgpm} is redefined as $T = \tanh [w
(t\pm t_0)]$.

\subsection{{\sf BiII} spiky features}
\label{BiII}

High velocity spike transitions $\Thi$ originate from Kasner points with $1 < \ue
<\infty$, i.e., from sector $(123)$; low velocity spike solutions $\Tlo$ from
points with $\textfrac{1}{2} < \ue < 1$, which covers merely the upper part
of sector $(132)$. The third type of spiky feature, which we call a
\textsf{BiII} spiky feature, fills the rest of sector $(132)$:

\begin{itemize}
\item[\textsf{BiII}] The limit of a $\textsf{BiII}$ spiky
    feature as $t\rightarrow -\infty$ is a Kasner point on
    sector (132) with $0 < \ue < \frac12$, i.e., $u > 2$.
    The trajectories $z\neq 0$ of a $\textsf{BiII}$ spiky
    feature coincide with the $\mathcal{T}_{N_1}$ transition
    that originate from the Kasner point $\ue$ on sector (132)
    and end in sector (312),
    while the spatial points of the
    spike surface $z = 0$ are left behind at the Kasner point $\ue$ on sector (132).
    Accordingly, the limit as $t\rightarrow \infty$
    is a generalized Kasner metric with discontinuous
    Kasner parameter such that
	 \begin{equation}\label{spikemapii}
	\ue_+(z) =
	\begin{cases}
	\ue_-  & z = 0 \,, \\[0.5ex]
	-\ue_- & z \neq 0\,.
	\end{cases}
	\end{equation}
	  In terms of the standard Kasner parameter we have
          $u_- \mapsto u_+ = u_-$
	  and $u_+ = u_- -1$, respectively.
\end{itemize}

A \textsf{BiII} spiky feature is intimately connected with the `second half'
of a low velocity spike solution $\Tlo$ in the neighborhood of $z=0$.
Specifically, consider a $\Tlo$ characterized by $w = 1
-2\ue$ with $\ue \in (0,\frac{1}{2})$, (inverse) coordinate spike width
$c_z$, and a time offset $t_0$, i.e., $t$ is replaced by $t + t_0$
in~\eqref{metricspike}. For a given value of $z=\tilde{z} \neq 0$, if $c_z$
is sufficiently small, the trajectory of $\tilde{z}$ will resemble the
concatenation of the low velocity spike surface trajectory $\mathcal{T}_S$
and a $\mathcal{T}_{N_1}$ transition originating from $\ue$. It is possible
to choose a (large) value of $t_0$, which is fine-tuned w.r.t.\ $c_z$, such
that the point $t = 0$ of the trajectory is close to the midpoint of the
$\mathcal{T}_{N_1}$ transition. This motivates taking the combined limit $c_z
\rightarrow 0$ and $t_0 \rightarrow \infty$ such that $2 c_z e^{(1-w) t_0} =
1$, which, via~\eqref{pe}, yields
\begin{subequations}\label{biIIpe}
\begin{align}
\pe_1 & = \frac{1}{3}(1 + \Sigma_1) =
\frac{-\ue \big(1 - (1 - 2 \ue)^2 z^2 e^{4 \ue t}\big)}%
{f(\ue) + f(-\ue) (1 - 2 \ue)^2 z^2 e^{4 \ue t}} \,,\\[0.5ex]
\pe_2 & = \frac{1}{3}(1 + \Sigma_2) =
\frac{2-(1-\ue) \big(1 - (1 - 2 \ue)^2 z^2 e^{4 \ue t}\big)}%
{f(\ue) + f(-\ue) (1 - 2 \ue)^2 z^2 e^{4 \ue t}}\,,
\\[0.5ex]
N_1 & = \frac{-12 \ue (1 - 2 \ue) z e^{2 \ue t}}%
{f(\ue) + f(-\ue) (1 - 2 \ue)^2 z^2 e^{4 \ue t}}\,,
\end{align}
\end{subequations}
while $R_3$ and $N_{1 2}$ are zero. The above expression is not a solution to
Einstein's equations, but it is an inhomogeneous solution of the equations on
the local boundary.

\section{Asymptotics and concatenation in the OT case}\label{sec:OT}

In the present section we restrict our attention to the OT subclass of the
$G_2$ models, of which the $T^3$ Gowdy models are a prominent special case,
see Appendix~\ref{app:topology}. The metric of the OT $G_2$ models is of the
form~\eqref{metricn}, where $n_2 = n_3 = 0$, i.e.,
\begin{equation}\label{G2ot}
ds^2 = - N^2(dx^0)^2 +
e^{-2b^1}\left(dx + n_1 dy\right)^2 + e^{-2b^2}dy^2 +
e^{-2b^3}dz^2\,,
\end{equation}
where $b^\alpha = b^\alpha(x^0,z)$ $\forall \alpha$ and $n_1=n_1(x^0,z)$. The
area time gauge entails $x^0 = t = b^1 + b^2$ (modulo a constant) and $N =
-\sqrt{\det {}^{(3)}\!g}$. Recall from section~\ref{sec:basics} that the
conformally Hubble-normalized rotation variable $R_1$ vanishes identically
under the OT assumption.

\subsection{BKL concatenation in OT compatible Bianchi models}
\label{BKLconcatOT}

Let us briefly recapitulate the basic facts about \textit{spatially
homogeneous} models in the present context. The existence of two commuting
Killing vector fields is incompatible with the defining properties of the
Bianchi type~VIII and~IX models. The Bianchi types that are compatible are
types~$\mathrm{VII}_h$ and~$\mathrm{VI}_h$, and the `lower types', which can
be obtained by means of Lie contractions. However, it is only the symmetry
adapted frames of Bianchi types~$\mathrm{VII}_0$, $\mathrm{VI}_0$, II, and~I
that are compatible with a line element~\eqref{G2ot} in the area time gauge,
cf.~\cite{elsetal02}.

\begin{figure}[ht]
\centering
\includegraphics[height=0.37\textwidth]{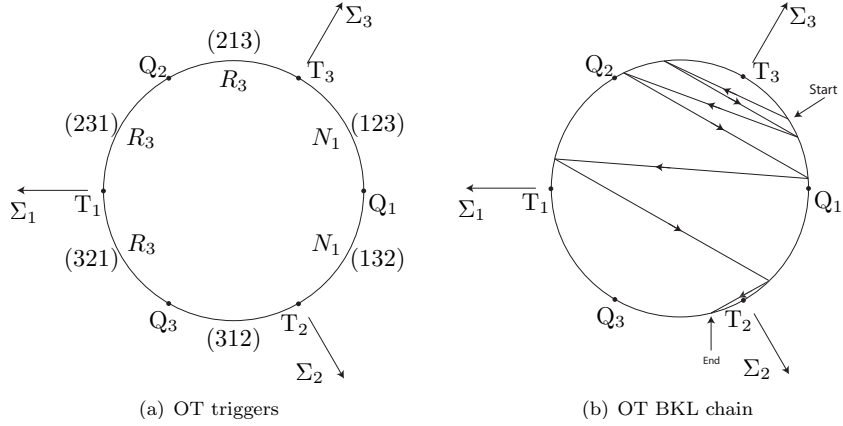}
        \caption{In the OT case each sector is associated with a `trigger' ($N_1$ or $R_3$) except
          for sector $(312)$ at which these variables are stable. As a consequence, OT BKL chains
          (i.e., concatenations of $\mathcal{T}_{N_1}$ and $\mathcal{T}_{R_3}$ transitions)
          are finite and terminate at sector $(312)$.}
           \label{OTtriggchain}
\end{figure}

The initial singularity of the spatially homogeneous OT $G_2$ vacuum models
is known to be a Kasner type singularity, i.e., the asymptotic behavior is
that of a Kasner solution. This can be understood in connection with
Fig.~\ref{OTtriggchain}. For Bianchi models of type~$\mathrm{VI}_0$ and
$\mathrm{VII}_0$, the non-trivial Hubble-normalized variables are the
diagonal shear variables and $N_1$ and $R_3$. From~\eqref{R13lin} and
Fig.~\ref{frametrans}(b), and~\eqref{N1lin} and Fig.~\ref{curvtrans},
respectively, we see that $R_3$ is an unstable variable (`triggered') in three
sectors of the Kasner circle while $N_1$ is triggered in two sectors, see
Fig.~\ref{OTtriggchain}(a). Sector (312), however, is a stable sector, which means
that $N_1$ and $R_3$ are stable variables, i.e., a solution with initial data
close to a Kasner point of sector (312) must converge to a Kasner point of that
sector, because $N_1$ and $R_3$ decay exponentially. It is thus intuitive
(and straightforward to prove, see, e.g.,~\cite{heietal09}) that every
type~$\mathrm{VI}_0$ or type~$\mathrm{VII}_0$ model converges to a fixed
point on the stable sector.

Considering Figs.~\ref{frametrans}(b) and~\ref{curvtrans} together, it is
clear that frame transitions and curvature transitions can be
\textit{concatenated} by identifying the `final' fixed point ($\omega$-limit
point) of one transition with the `initial' fixed point ($\alpha$-limit
point) of another transition, see Fig.~\ref{OTtriggchain}(b). Concatenation of
$\mathcal{T}_{R_3}$ and $\mathcal{T}_{N_1}$ transitions yields finite
heteroclinic chains,\footnote{A heteroclinic orbit is an orbit (i.e. solution
trajectory) that starts ($\alpha$-limit) and ends ($\omega$-limit) at two
different fixed points. A heteroclinic chain is a sequence of heteroclinic
orbits such that the $\omega$-limit point of one orbit is the $\alpha$-limit
point of the subsequent orbit.} which we refer to as OT \textit{BKL chains};
these chains terminate on the stable sector $(312)$.

The relevance of the OT BKL chains is immediate: The evolution of initial
data sufficiently close to a Kasner point of any other than the stable sector
sector (or: the evolution of data sufficiently close to a transition) is
approximately described an OT BKL chain, which yields a finite number of
`oscillations' between Kasner states, mediated by $\mathcal{T}_{N_1}$ and
$\mathcal{T}_{R_3}$ transitions, and convergence to a Kasner point on the
stable sector $(312)$. However, it should be pointed out that the evolution
of initial data far from the Bianchi type~I and the type~II subsets (i.e.,
far from the Kasner circle and the transition orbits) need not bear any
relation to OT BKL chains, although there is still convergence to a Kasner
point on the stable sector. For a comprehensive study of
type~$\mathrm{VI}_0$/$\mathrm{VII}_0$ dynamics we refer to~\cite{heirin09,
heiugg09a} and references therein.

\subsection{Asymptotics of spatially inhomogeneous OT models}
\label{spikeorbits}

According to the BKL conjecture, asymptotic locality means that in the
\textit{inhomogeneous} case there exists a generic set of timelines whose
evolution is governed asymptotically by the flow on the local boundary (which
entails that the evolution of a single timeline is congruent with the
evolution of a spatially homogeneous model.) This does \emph{not} mean,
however, that the dynamics of spatially homogeneous models is sufficient to
describe the dynamics of inhomogeneous models. On the contrary, spatially
inhomogeneous OT models behave quite delicately in general.

However, Fig.~\ref{OTtriggchain}(a) suggests a straightforward behavior for at least
certain classes of initial data. Consider, e.g., initial data that constitute
a small perturbation of a generalized Kasner metric associated with the
stable sector (312) on $\mathrm{K}^{\ocircle}$, i.e., $\Sigma_1, \Sigma_2,
\Sigma_3$ are functions of the spatial variable $z$ with values close to
sector (312), while the remaining variables and the Hubble-normalized spatial
frame derivatives are small; note that we do not exclude that, e.g., $N_1$
has zeros. Since neither $N_1$ nor $R_3$ are triggered but decay
exponentially, we expect convergence to a generalized Kasner solution on
sector (312). This is an asymptotic state that is often termed
`asymptotically velocity dominated'; we refer to Appendix~\ref{app:OT}, where
we discuss results that turn the intuitive picture into a rigorous one in the
case of $T^3$ Gowdy models.

Prescribing `almost Kasner' initial data close to sector (321), on the other
hand, leads to an exponential decay of $N_1$, while $R_3$ is triggered, so
that an (approximate) $\mathcal{T}_{R_3}$ frame transition ensues at each
spatial point,\footnote{Technical difficulties occur if $R_3$ is zero
initially at some value of $z$; in this case a `false spike' ensues. We
refrain from discussing this case since it is a gauge effect (a ``nuisance''
as stated in~\cite{rin06}) and focus on `true' spatial structures instead. We
note however that a complete description of a generic singularity in an
Iwasawa frame necessarily involves gauge features of this kind.} which in
turn leads to the convergence of the solution to a generalized Kasner metric
of sector (312), as before.

Likewise, initial data close to sectors (231) or (213) lead to an initial
frame transition $\mathcal{T}_{R_3}$, which takes the solution to sectors
(132) or (123) where $N_1$ is the trigger. If $|N_1| > 0$ (in the optimal
case: uniformly) we are led to expect the familiar behavior of the spatially
homogeneous type~$\mathrm{VI}_0$/$\mathrm{VII}_0$ case, i.e., a subsequent
(approximate) $\mathcal{T}_{N_1}$ curvature transition at each spatial point;
accordingly, each timelike follows an (approximate) OT BKL chain. On the
other hand, \textit{if $N_1$ has zeros} for certain values of the spatial
coordinate $z$, the simple picture inspired by spatially homogeneous dynamics
fails. At these spatial surfaces it cannot be the growth of $N_1$ that drives
the further evolution of the solution. Instead, spatial gradients take the
role of triggers at these special values of $z$ (and their neighborhoods) and
spatial structures (`spikes') develop (which are `true' spikes in the sense
of being gauge invariant features). These spikes are approximately described
by the explicit spike solutions $\Thi$, $\Tlo$, and the $\textsf{BiII}$ spiky
feature in a sense that we discuss in section~\ref{spikeconOT}.

We emphasize that the evolution of general initial data is rather complicated
(and the heuristic considerations fail). However, asymptotic velocity
dominance remains true in general, i.e., solutions converge to a generalized
Kasner metric (which is in general associated with a discontinuous Kasner
parameter); see Appendix~\ref{app:OT}.

\subsection{Spike concatenation in OT models}
\label{spikeconOT}

\begin{figure}[ht]
\centering
\includegraphics[width=0.86\textwidth]{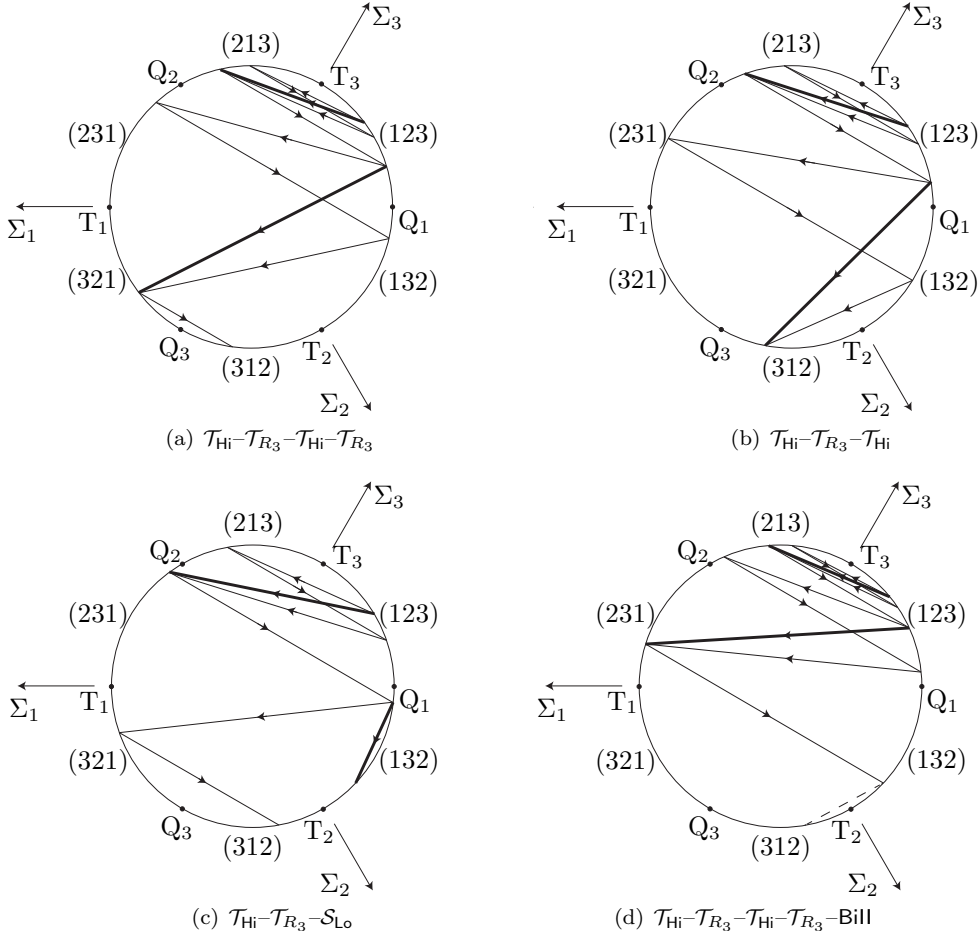}
        \caption{OT spike chains in the projection onto
        $(\Sigma_1, \Sigma_2, \Sigma_3)$-space; depicted are chains of the spike surface trajectories ($z = 0$)
        and trajectories associated with $|z| \gg 1$ (which are OT BKL chains). An OT spike chain
        is a high velocity chain (i.e., a concatenation
        of $\Thi$ transitions and $\mathcal{T}_{R_3}$ frame transitions)
        ending with a $\mathcal{T}_{R_3}$ transition (a) or a $\Thi$ transition (b);
        or there is one additional (final) transition: a low velocity solution $\Tlo$ (c) or
        a \textsf{BiII} spiky feature (d). The latter is depicted by a dashed line.}
        \label{hlconc}
\end{figure}

In section~\ref{BKLconcatOT} we have concatenated curvature transitions and
frame transitions to build OT BKL chains. Analogously, high velocity spike
transitions $\Thi$ and frame transitions $\mathcal{T}_{R_3}$ can be
concatenated by identifying the `final' Kasner point of one transition with
the `initial' Kasner point of another transition, see Figs.~\ref{hlconc}(a)
and~\ref{hlconc}(b). Note that this is possible because the entire family of
curves representing a $\Thi$ transition converges to a point on
$\mathrm{K}^\ocircle$ as $t\rightarrow -\infty$ and to another point on
$\mathrm{K}^\ocircle$ as $t\rightarrow \infty$. Concatenation yields
finite\footnote{Concatenation in the opposite direction of the time
coordinate, i.e., away from the singularity, yields infinite chains
converging to the Taub point $T_3$.} chains, which we refer to as
\textit{high velocity chains}.

\begin{figure}[ht]
\centering{
        \includegraphics[height=0.37\textwidth]{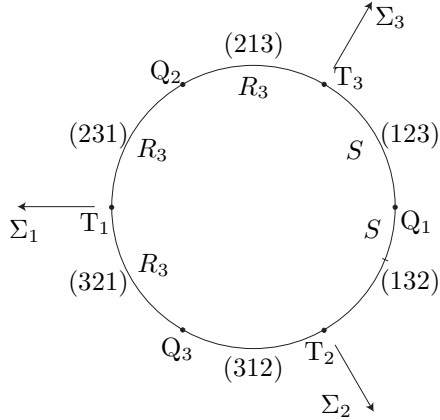}}
        \caption{`Triggers' for the evolution of the spike surface $z = 0$ in the OT case.
          In sectors $(213)$, $(231)$, and $(321)$, $R_3$ triggers frame transitions $\mathcal{T}_{R_3}$.
          Since $N_1 = 0$ at $z = 0$, $N_1$ does not exist as a trigger; instead spatial
          gradients (`$S$') take the role as triggers and lead to a spike surface trajectory
          $\mathcal{T}_S$, see Fig.~\ref{spiketrans}. These triggers cover sector $(312)$ and
          $(132)$ with $u>2$. Sector $(312)$ and sector $(132)$ with $u < 2$
          are stable in this context.}
        \label{OTSpiketrigg}
\end{figure}

Tracking the extended Kasner parameter along a high velocity chain we find an
alternation between the $\Thi$ law $\ue\mapsto 1 - \ue$ and the
$\mathcal{T}_{R_3}$ law $\ue \mapsto -(\ue + 1)$, whose combination results
in $\ue \mapsto \ue - 2$. A high velocity chain either ends at a point on the
stable Kasner sector (312), where the final transition can be a
$\mathcal{T}_{R_3}$ as in Fig.~\ref{hlconc}(a) or a $\Thi$ as in
Fig.~\ref{hlconc}(b), or the chain ends at a point on sector (132). The
latter case is particularly interesting, since the high velocity chain can
then be continued by a $\Tlo$ solution or a $\textsf{BiII}$ spiky feature,
depending on its location in sector $(132)$.  We use the term `\textit{OT
spike chains}' as an umbrella term for the two cases, i.e., an OT spike chain
is either a high velocity chain ending in sector (312) or a high velocity
chain followed by a $\Tlo$ solution or a $\textsf{BiII}$ spiky feature.
Note that the continuation of OT spike chains beyond this point is impossible
(because $\mathcal{T}_{R_1}$ transitions do not exist in the OT context; see,
however, the next section). Note that for an OT spike chain ending with a
$\Tlo$ solution or a $\textsf{BiII}$ spiky feature the trajectories of the
points on the spike surface $z = 0$ end at sector $(132)$ with $0<\ue<1/2$,
while points with $z \neq 0$ are transported to the stable sector $(312)$.
Therefore, in a vague sense, the part of sector $(132)$ with $0<\ue<1/2$ is
`stable' for the spike surface, see Fig.~\ref{OTSpiketrigg}.

The role of OT BKL chains for Bianchi models of type~$\mathrm{VI}_0$
and~$\mathrm{VII}_0$ has been discussed in section~\ref{BKLconcatOT}.
Numerical evidence~\cite{limetal09} suggests that OT spike chains fill the
same function for inhomogeneous solutions with spikes. In the following we
give a detailed description. For simplicity, we restrict our attention to
\textit{symmetric} initial data sets, by which we mean initial data such that
$\Sigma_1$, $\Sigma_2$, $\Sigma_3$, and $R_3$ are even functions in $z$,
while $N_1$ and $N_{1 2}$ are odd functions (where we are interested in the
non-degenerate case for which the gradients at $z = 0$ are non-vanishing).%
\footnote{In addition we assume the `genericity condition' that $R_3\neq 0$
at the spike surface $z= 0$.} This type of initial data fixes the location of
the spike surface at $z = 0$.\footnote{In general, this type of initial data
fixes the location of \textit{a} spike surface at $z = 0$. During the
evolution of the data, it is possible that other (in general `moving') spikes
form at different values of $z$. We refer to the concluding remarks.} Note
that the explicit spike solutions $\Thi$, $\Tlo$, and the $\textsf{BiII}$ spiky
feature, see~\eqref{pe} and~\eqref{biIIpe}, respectively, are of this
symmetric type. Since the location of the spike surface is fixed, it is
simple to calculate the particle horizons associated with the spike world
sheet $z = 0$. Since a null vector orthogonal to the surfaces of symmetry is
of the form $(1,0,0, \pm e^{-t})^{\mathrm{T}}$, see~\eqref{G2ot}, the
particle horizon at $t$ of the $z= 0$ timelines is $[-e^{-t}, e^{-t}]$ times
the symmetry surfaces; in particular, the particle horizons are shrinking
rapidly (in $z$-direction) as $t\rightarrow \infty$. Hence the dynamics of a
solution in a neighborhood of the spike world sheet for $t \geq t_0$ is
completely determined by the initial data of the solution in a neighborhood
of $[-e^{-t_0}, e^{-t_0}]$. In particular, changing the initial data for an
explicit spike solution outside a neighborhood of $[-e^{-t_0}, e^{-t_0}]$
does not affect the evolution~\eqref{metricspike} of the spike surface and
its immediate neighbors.\footnote{The asymptotic dynamics is thus local in
the sense that global topological aspects are irrelevant once the particle
horizon scale has become sufficiently small compared to the `global scale'
associated with the topology.} In the following we restrict our attention to
(symmetric) initial data on a neighborhood of $[-e^{-t_0}, e^{-t_0}]$ and
consider the evolution of that data on its domain of dependence.

A solution $S(t,z)$ with initial data sufficiently close (in, e.g., a
$\mathcal{C}^{k,\alpha}$ norm) to an explicit spike transition $\Thi$ will
remain close to that solution for some time interval; temporarily we may thus
speak of an \textit{approximate spike transition}. The solution $S(t,z)$ is
approximated by $\Thi$ at least up to a point when the two solutions reach a
neighborhood of the Kasner circle $\mathrm{K}^\ocircle$. While $\Thi$
continues to converge to a fixed point on $\mathrm{K}^\ocircle$, the solution
$S(t,z)$ will, in general, stray off course---it ceases to be an approximate spike
transition. The initially small deviation of $S(t,z)$ from $\Thi$ amplifies
(in the sense of increasing relative errors) and leads to different behavior
in the variable $R_3$. In contrast to $\Thi$, this variable is triggered at
$\mathrm{K}^\ocircle$ and $S(t,z)$ undergoes an approximate
$\mathcal{T}_{R_3}$ frame transition. This behavior can be understood,
qualitatively and quantitatively, by considering a linearization of the
equations at $\mathrm{K}^\ocircle$, but since such an analysis goes beyond
the scope of the present paper it will be
pursued elsewhere. 
For the present purposes we restrict ourselves to referring to the strong
numerical evidence we have obtained for the described behavior: The numerics
suggest that the solution $S(t,z)$ shadows an OT spike chain
\begin{equation}
\Thi \longrightarrow \mathcal{T}_{R_3} \longrightarrow \dotsb \longrightarrow \Thi
\longrightarrow \mathcal{T}_{R_3}
\longrightarrow \begin{cases} \Thi \\ \,- \\ \Tlo \\ $\textsf{BiII}$ \end{cases}\,,
\end{equation}
and thus $S(t,z)$ is an \textit{approximate OT spike chain}. Furthermore, the
numerical evidence shows that an approximate OT spike chain $\mathcal{S}$
shadows an exact OT spike chain with an increasing degree of accuracy; in
particular, the asymptotic limit of $S(t,z)$ is a small perturbation of the
limit of the OT spike chain itself.

For an approximate spike transition we \textit{define} its
(inverse) spike width $c_z$ through~\eqref{spikewidth}, i.e.,
\begin{equation}\label{widthdef}
c_z = \frac{1 + 2 \pe_1|_{z=0}}{24 \pe_1|_{z=0}}\,
\big(\ue_- + f(\ue_-) \pe_1|_{z=0}\big)^{-\ue_-/w}
\big( {-\ue_+} - f(\ue_+) \pe_1|_{z=0}\big)^{\ue_+/w}
\, \partial_z N_1\,\big|_{z = 0} \,,
\end{equation}
where we recall that $\pe_1 = \textfrac{1}{3} (1+\Sigma_1)$. The inverse
width $c_z$ of an approximate spike transition will in general not be a
constant exactly, but depend on when~\eqref{widthdef} is evaluated. However,
for an approximate spike transition, $c_z$ will be approximately constant, at
least up to the point where the deviation from $\Thi$ becomes too large. Note
that definition~\eqref{widthdef} is not unique in the sense that it can be
replaced by similar definitions involving, e.g., $N_{1 2}$ and $\pe_2$
instead of $N_1$ and $\pe_1$. Evidently, for exact spike transitions this
does not cause any difference, while for approximate spike transitions the
resulting value will indeed differ from~\eqref{widthdef} in general. However,
the slight ambiguities quickly converge to zero when the approximate spike
transitions are approximated by the exact transitions with an increasing
degree of accuracy.

Considering an approximate OT spike chain $S(t,z)$ a natural question to ask
concerns the behavior of the evolution of the spike width. Suppose that
$S(t,z)$ is approximated for some time interval by an exact spike transition
$\Thi$ with parameter $\ue_0$, inverse width $c_0$, and time offset ${-t_0}$,
i.e., by~\eqref{metricspike} with the replacements $z \mapsto c_0 z$ and
$t\mapsto t - t_0$. In a (much) later time interval, the solution $S(t,z)$ is
then approximated by the subsequent $\Thi$ of the chain; we have $\ue_1 =
\ue_0 - 2$, inverse coordinate width $c_1$, and time offset $t_1$ (where $t_1
\gg t_0$). In Appendix~\ref{spikewidthevol} we show that
\begin{equation}\label{spikeevol}
c_1  = e^{t_1 -t_0} \,c_0 = e^{\Delta t} \, c_0\,.
\end{equation}
The time $\Delta t$ between two subsequent (approximate) spike transitions is
dominated by the time the solution spends in the neighborhood of
$\mathrm{K}^\ocircle$. The closer the solution $S(t,z)$ is to an exact OT
spike chain, the greater $\Delta t$ will be. As a consequence the coordinate
width of the spike decreases rapidly toward the singularity, since the
inverse width increases rapidly.

\section{Concatenation in the general $G_2$ case}\label{sec:genG2}

\begin{figure}[ht]
\centering
\includegraphics[width=0.85\textwidth]{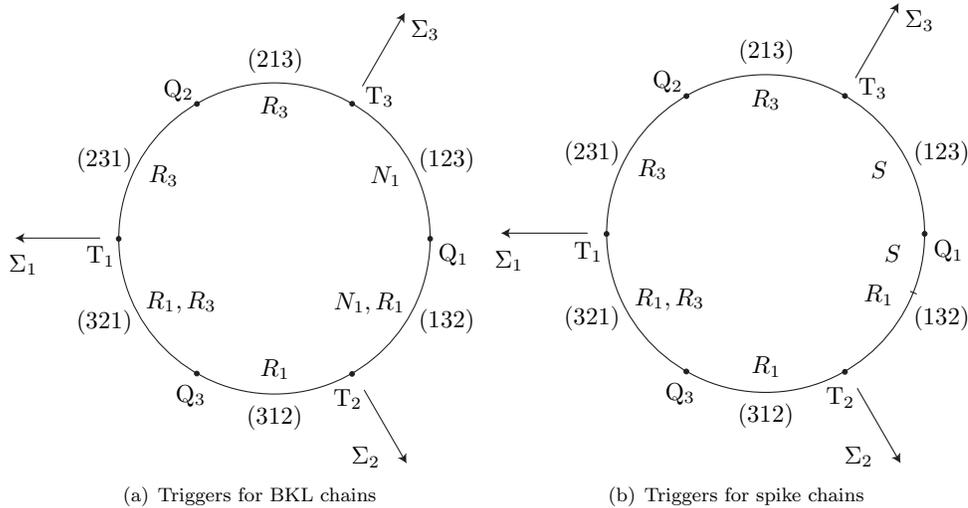}
  \caption{Triggers for general $G_2$ models.
  Subfigure (a) and (b) generalizes Fig.~\ref{OTtriggchain}(a) and Fig.~\ref{OTSpiketrigg}, respectively.}
  \label{BKLtriggers}
\end{figure}

In contrast to the OT case, the Hubble-normalized frame variable $R_1$ does
not vanish identically for the general $G_2$ models. As a consequence, sector
(312), which is stable for the OT models, see Fig.~\ref{OTtriggchain}(a), possesses
an unstable manifold which is associated with the variable $R_1$, see
Fig.~\ref{BKLtriggers}. The same is true for the part of sector $(132)$ with
$0<\ue<1/2$, which is `stable' for the spike surface trajectories in the OT
context, see Fig.~\ref{OTSpiketrigg}, but unstable through
$\mathcal{T}_{R_1}$ frame transitions for general $G_2$ models, see
Fig.~\ref{BKLtriggers}(b). Intuitively it follows that the instability of the
entire Kasner circle (except for, possibly, the Taub points) prevents
convergence of (generic) solutions to generalized Kasner solutions and leads
to \textit{oscillatory singularities}. In this section we present analytical
and numerical evidence that corroborate this conjecture.
%

\subsection{BKL concatenation and BKL chains}
\label{BKLconcat}

\begin{figure}[ht]
 \centering{
  \includegraphics[height=0.45\textwidth]{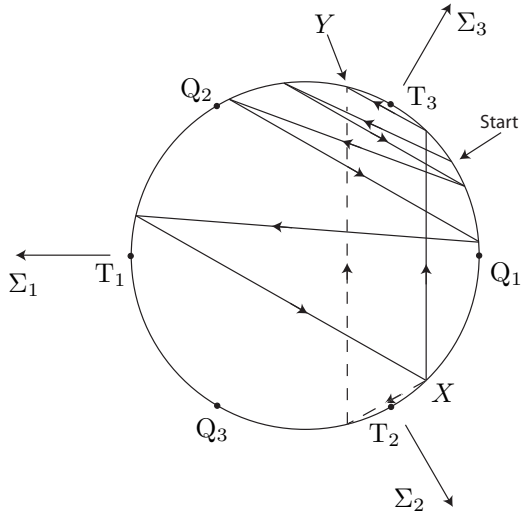}}
\caption{A BKL chain is an (in general) infinite heteroclinic chain consisting
of $\mathcal{T}_{N_1}$, $\mathcal{T}_{R_1}$ and
$\mathcal{T}_{R_3}$ curvature and frame transitions.
The paths are not unique; e.g., at the point ${X}$ two continuations are possible,
a frame transition $\mathcal{T}_{R_1}$ or a curvature transition $\mathcal{T}_{N_1}$ (dashed line).
However, the different paths rejoin at the point ${Y}$ after an additional curvature or frame transition.
At the point ${Y}$ the chain continues with a (not shown) $\mathcal{T}_{R_3}$ transition.}
\label{BKLSequence}
\end{figure}

Concatenation of frame transitions $\mathcal{T}_{R_1}$, $\mathcal{T}_{R_3}$,
and curvature transitions $\mathcal{T}_{N_1}$ yields heteroclinic chains,
which we refer to as a `\emph{BKL chains}', see Fig.~\ref{BKLSequence}. As
opposed to OT chains, BKL chains are \textit{infinite} in general.

We note that a BKL chain is not uniquely determined by an (initial) fixed
point on $\mathrm{K}^{\ocircle}$.\footnote{Note that this is different from
the Mixmaster case associated with Bianchi types~VIII and~IX. Recall that the
symmetry adapted frames of types~VIII and~IX are not of the Iwasawa type;
instead it is strongly preferable to use a symmetry group compatible Fermi
propagated orthonormal frame, which yields three types of curvature
transitions associated with the three spatial directions, while frame
transitions are absent. The associated BKL chains do not exhibit the
ambiguities of Fig.~\ref{BKLSequence}; see, e.g.,~\cite{heiugg09b} for
details.} This is because there are two sectors with two triggers (i.e.,
unstable modes) instead of one, see Fig.~\ref{BKLtriggers}: In sector (132)
it is possible to continue the chain with a $\mathcal{T}_{N_1}$ curvature
transition or a $\mathcal{T}_{R_1}$ frame transition; in sector (321) it is
possible to continue with a $\mathcal{T}_{R_1}$ or a $\mathcal{T}_{R_3}$
frame transition. Note, however, that different paths rejoin after two or
three transitions, see Fig.~\ref{BKLSequence}; in any case, the scenario of
two triggers at a fixed point requires closer attention; we refer to the
discussion below.

In analogy with the role of OT BKL chains in the OT context, the relevance of
BKL chains in the present context is twofold. On the one hand, the role of
BKL chains for spatial homogeneous models is immediate: Since the
symmetry-adapted frames of general Bianchi type $\mathrm{VI}_{-1/9}$ models
are Iwasawa frames, these models are of the form~\eqref{metricn} where the
metric is manifestly spatially homogeneous, see~\cite{heietal09}. The
asymptotic evolution toward the initial singularity of general type
$\mathrm{VI}_{-1/9}$ models is expected to be represented by\footnote{The
exact meaning of `being represented by' is a delicate point. Clearly, the
evolution of initial data close to a Kasner point is approximately described,
at least for a finite time, by a finite piece of a BKL chain (`finite
shadowing'). However, the study of Mixmaster dynamics~\cite{reitru10, beg10,
lieetal10} suggests the possibility that a generic solution asymptotically
follows `its' BKL chain forever.} BKL chains which lead to infinite
oscillations between different Kasner states; the singularity can therefore
be referred to as being \textit{oscillatory}. On the other hand, there exists
heuristic evidence~\cite{heietal09} and numerical
support~\cite{andetal05,gar04} that the asymptotic evolution towards a
generic spacelike singularity along a generic (open) set of timelines of a
general inhomogeneous vacuum (or `vacuum dominated') solution of the Einstein
equations (with $|N_1R_1R_3| > 0$) is represented by BKL chains as well. (For
the perfect fluid case and `vacuum dominance' we refer to~\cite{sanugg10} and
references therein.)

Asymptotic shadowing necessarily brings solutions (i.e.,
type~$\mathrm{VI}_{-1/9}$ solutions or single timelines of inhomogeneous
solutions) to a neighborhood of a Kasner point where two variables are
`triggered' (i.e., grow exponentially). Consider, e.g., the case of a
trajectory following a $\mathcal{T}_{R_3}$ transition to a fixed point ${X}$
on sector (132) of $\mathrm{K}^\ocircle$, where $N_1$ and $R_1$ are triggers,
see Figs.~\ref{BKLtriggers} and~\ref{BKLSequence}. If $|N_1| \ll |R_1|$ ($
\ll 1$) when the trajectory enters a neighborhood of the fixed point ${X}$,
$R_1$ will be triggered and a $\mathcal{T}_{R_1}$ will follow; if $|R_1| \ll
|N_1|$, $N_1$ will be triggered and a $\mathcal{T}_{N_1}$ will follow.
Clearly, it is possible to prescribe initial data such that both $R_1$ and
$N_1$ grow to macroscopic size, which means that a so-called multiple
transition (here: a mixed curvature/frame transition) follows. Generic BKL
chains enter a neighborhood of the fixed point ${X}$ infinitely many times,
and generic solution trajectories are thus expected to do likewise. However,
the analysis of~\cite{heietal09} suggests that the multiple transition
scenario does not occur infinitely often; generically, in the asymptotic
limit, trajectories follow single transitions (and thus BKL chains), a result
that is consistent with the billiard picture of Damour~{\it et
al}~\cite{dametal03}. The question of which of the two possible transitions,
$\mathcal{T}_{R_1}$ or $\mathcal{T}_{N_1}$, is followed more frequently, is
delicate. Linearization the system at ${X}$ yields the eigenvalues associated
with the variables $R_1$ and $N_1$. If the eigenvalue associated with, say,
$R_1$ is considerably larger than the other, triggering a $\mathcal{T}_{R_1}$
transition seems more likely than a $\mathcal{T}_{N_1}$ transition. It is
possible, however, that $R_1$ is almost always sufficiently small against
$N_1$ when the trajectory comes close to ${X}$, which would make
$\mathcal{T}_{N_1}$ transitions more frequent than expected. In any case, the
fact that different paths rejoin, makes the ambiguity in paths a transient
problem.

\subsection{Spike concatenation in the general $\bm{G_2}$ models}
\label{concatG2}

While the oscillatory asymptotic dynamics of generic timelines is represented
by BKL chains, the evolution of spike timelines, and their neighbors, bears
no relation to spatially homogeneous dynamics; the BKL picture breaks down.
This is because $N_1$ is zero at the spike surface which results in a loss of
the trigger $N_1$ in sectors $(123)$ and $(132)$; in other words,
$\mathcal{T}_{N_1}$ curvature transitions cannot take place. However, from
Fig.~\ref{BKLtriggers}(b) we see that the general $G_2$ case has at least one
trigger at each sector of $\mathrm{K}^\ocircle$ (which is due to the existence 
of a non-zero $R_1$), where spatial gradients take the
role as triggers in sector $(123)$ and part of sector $(132)$. We will
argue that this `instability' of the Kasner circle results in oscillatory
dynamics which is represented by infinite \textit{spike chains}.

As in the case of OT spike chains, see section~\ref{spikeconOT}, high velocity
spike transitions $\Thi$ and \mbox{$\mathcal{T}_{R_1}$ \& $\mathcal{T}_{R_3}$} frame
transitions can be concatenated straightforwardly since the entire family of
curves re\-presenting a $\Thi$ converges to a point on $\mathrm{K}^\ocircle$
as $t\rightarrow -\infty$ and to another point on $\mathrm{K}^\ocircle$ as
$t\rightarrow \infty$. However, the situation for low velocity spike
solutions $\Tlo$ is different, because the limit on $\mathrm{K}^\ocircle$
of the trajectory of the spike surface $z = 0$ differs from the 
(common) limit of the $z\neq 0$ timelines, see Figs.~\ref{highandlowspiketranstime}(d) and~\ref{Lowvel}.
It is obvious that low velocity spike solutions $\Tlo$ do not immediately
fit into the network of transitions.

However, we will see that low velocity spike solutions naturally
\textit{combine} with a frame rotation in $R_1$ and part of a high velocity
transition to form a \textit{joint low/high velocity spike transition} which
we denote by $\Tco$. This spike transition is characterized by the same 
fundamental property as $\Thi$ transitions that the entire family of curves 
representing $\Tco$ emerges from one point and 
ends at another point on $\mathrm{K}^\ocircle$. 
There is thus a stark contrast between the OT case and the general $G_2$ case.
In the OT case, a $\Tlo$ solution and perturbations thereof develop `permanent'
spatial structures since the timelines of the spike surface and the 
non-spike timelines converge to two different Kasner points, see Figs.~\ref{highandlowspiketranstime}(d)
and~\ref{Lowvel}; in the general $G_2$ case, these spatial structures 
form \textit{and `un-form'} since $\Tlo$ is merely the first part of
the transition $\Tco$: The timelines rejoin eventually. 
Spikes are transient features.
We will see that the initial Kasner point and the final Kasner point
of a joint low/high velocity spike transition $\Tco$ are
are related by the spike map~\eqref{spikemap}, which thus
characterizes \textit{both} $\Thi$ and $\Tco$ transitions; we will argue that
this property is not coincidental.

In order to unveil the structures leading to joint low/high velocity
spike transitions $\Tco$ we use information about \emph{timing}.
First, consider Fig.~\ref{highandlowspiketranstime}(d) and 
compare the trajectory of the spike surface $z = 0$ with the
trajectories associated with $z\neq 0$ of a neighborhood $U_0$ of $z= 0$ of
$\Tlo$. The (dashed) lines of constant
time show that the respective final Kasner points, $A_0$ and $A_{1}$, are
approached `simultaneously' (which is in contrast to the approach to the
initial Kasner state); specifically, let $X(t,z)$ denote the family of curves
representing the $\Tlo$ transition; then $|X(t,0) - A_0|$ and $|X(t,1)-A_1|$
are of the same order as $t\rightarrow \infty$. Furthermore, for $t \gg 1$,
the set $\{X(t,z) | \,z \in U_0\}$ is approximated by the straight line $S_A$
connecting the two Kasner points $A_0$ and $A_{1}$, which corresponds to the
$\mathcal{T}_{N_1}$ transition that originates from $A_0$, see
Fig.~\ref{Lowvel}.

Second, consider Fig.~\ref{highandlowspiketranstime}(a) or~\ref{highandlowspiketranstime}(b); let $Y(t,z)$ denote
the family of curves representing the $\Thi$ transition. The convergence as
$t\rightarrow -\infty$ to the initial Kasner point $B_0$ is non-uniform; the
(dashed) lines of constant time show that for $t \ll -1$, the set $\{Y(t,z) |
\,z \in \mathbb{R}\}$ is approximated by the straight line $S_B$ representing
the $\mathcal{T}_{N_1}$ transition connecting the initial Kasner point $B_0$
with the Kasner point $B_1$.

In the general $G_2$ case the variable $R_1$ does not vanish identically.
Consider initial data generated by a small perturbation of $\Tlo$ initial
data such that $|R_1| \neq 0$ is sufficiently small. (In accordance with the
considerations of section~\ref{spikeconOT} we assume symmetric initial data
sets; in particular, $R_1$ is an even function of $z$.) The associated
solution $S(t,z)$ shadows the $\Tlo$ transition closely; hence, for a large
$t_A$, $\{S(t_A,z) | \,z \} \approx S_A$. However, $S(t,z)$ cannot converge
to the Kasner circle since $R_1$ increases in a neighborhood of $S_A$,
cf.~Fig.~\ref{BKLtriggers}. Eventually, despite the initial smallness of
$R_1$, the solution $S(t,z)$ will be transported away from $S_A$ by a `frame
transition'.

\begin{figure}[ht]
\centering 
        \includegraphics[height=0.37\textwidth]{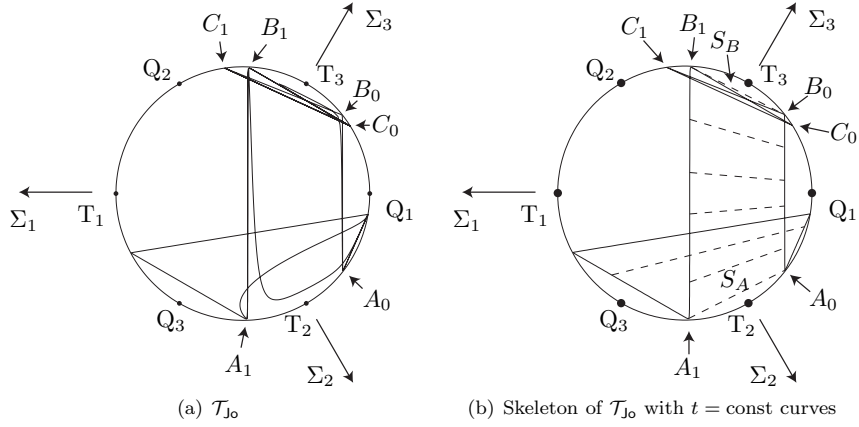}
        \caption{A joint low/high velocity spike transition $\Tco$ consists of a low velocity spike solution $\Tlo$ followed
          by a family of $\mathcal{T}_{R_1}$ frame transitions and the `second part' of a high velocity spike transition $\Thi$.
          In (a) we depict the numerical evolution of four timelines associated with different values of $z$ from $z = 0$ (spike 
          surface trajectory) to $|z| \gg 1$ (BKL chain). Subfigure (b) shows the curves of constant time (dashed lines) for a $\Tco$,
          which correspond to segments of the straight lines representing $\mathcal{T}_{N_1}$ transitions.}
        \label{G2hlconc2}
\end{figure}

A trajectory $\{S(t, \tilde{z}) | \,t\}$ with sufficiently large
$|\tilde{z}|$ (i.e., sufficiently far from the spike surface) is approximated
by a BKL chain: While $S(t,\tilde{z})$ approaches the Kasner point $A_1$,
$R_1$ grows steadily and 
a $\mathcal{T}_{R_1}$ frame transition follows, i.e., $S(t,\tilde{z})
\approx \mathcal{T}_{R_1}$ for some time interval, which brings
$S(t,\tilde{z})$ into the vicinity of the Kasner point $B_1$, where $R_1$
decreases rapidly; from there a $\mathcal{T}_{R_3}$ frame transition ensues
taking $S(t,\tilde{z})$ to the Kasner point $C_0$ where the trigger $N_1$
induces a $\mathcal{T}_{N_1}$ curvature transition that eventually
takes $S(t,\tilde{z})$ to a Kasner point $C_1$. For the trajectory $\{S(t,0)
|, t\}$ representing the spike surface, the scenario is straightforward as
well: $S(t,0)$ approaches the Kasner point $A_0$ while $R_1$ grows steadily;
a $\mathcal{T}_{R_1}$ frame transition follows, which brings $S(t,0)$ into
the vicinity of the Kasner point $B_0$ (where $R_1$ decreases rapidly); a
spike surface trajectory $\mathcal{T}_S$ follows and the
solution approaches the Kasner point $C_1$; in particular we observe that $S(t,0)$
\textit{rejoins} $S(t,\tilde{z})$ at $C_1$; we refer to Fig.~\ref{G2hlconc2}. The
described scenario is supported by heuristic arguments invoking
Fig.~\ref{BKLtriggers}(b) and the symmetry of the initial data set, and by
numerical simulations, exemplified in Figs.~\ref{G2hlconc2}(a)
and~\ref{G2hlconc2}(b).

Numerical experiments provide further information: For $t
> t_A$, the solution $S(t,z)$ is transported from $S_A$ to $S_B$, i.e., there
is $t_B > t_A$ such that $\{S(t_B, z)|\,z\} \approx S_B$, see
Figs.~\ref{G2hlconc2}(a) and~\ref{G2hlconc2}(b). Note that $R_1$ is
decreasing rapidly as the solution approaches $S_B$. The `transition' from
$S_A$ to $S_B$ occurs in a special manner: Curves of constant time, i.e.,
$\{S(t, z)|\, z\}$, $t_A < t < t_B$, see the dashed lines of
Fig.~\ref{G2hlconc2}(b), are straight lines corresponding to segments of the
paths of curvature transitions. The `frame transition' that maps $S(t_A, z)$
to $S(t_B,z)$ is therefore the (nonlinear) superposition of two motions: The
motion represented by the `ladder' of constant time slices of
Fig.~\ref{G2hlconc2}(b) and the motion of spatial points along $t =
\mathrm{const}$ lines. For $t$ close to $t_A$,~\eqref{pe1ag} yields
\begin{subequations}
\begin{equation}\label{p1A}
\tilde{p}_1 = \frac{-\ue_A \big( 1 - k_A^2 e^{4 \ue_A (t-t_A)} z^2\big)}{f(\ue_A) +
f(-\ue_A) k_A^2 e^{4 \ue_A (t-t_A)} z^2}\:,
\end{equation}
where $k_A$ is a constant and $\ue_A$ is associated with the Kasner point
$A_0$;~\eqref{p1A} describes the motion of spatial points $z$ along $S_A$ for
$t$ close to $t_A$. (Note the intimate connection of~\eqref{p1A} with a
\textsf{BiII} spiky feature, see section~\ref{BiII}.) For $t$ close to $t_B$,
on the other hand, we have
\begin{equation}\label{p1B}
\tilde{p}_1 = \frac{-\ue_B \big( 1 -k_B^2 e^{4 \ue_B (t-t_B)} z^2\big)}{f(\ue_B) +
f(-\ue_B) k_B^2 e^{4 \ue_B (t-t_B)} z^2}
= \frac{-\ue_A \big( 1 -k_B^2 e^{4 \ue_B (t-t_B)} z^2\big)}{f(\ue_A) +
f(-\ue_A) k_B^2 e^{4 \ue_B (t-t_B)} z^2} \:,
\end{equation}
\end{subequations}
where $k_B$ is another constant and $\ue_B$ is associated with $B_0$, i.e.,
$\ue_B = \ue_A^{-1}$ (because $\ue_B$ arises from $\ue_A$ through a
$\mathcal{T}_{R_1}$ transition). Equation~\eqref{p1B} describes the motion of
spatial points along $S_B$ when $t$ is close to $t_B$. Interpolation between
the two motions~\eqref{p1A} and~\eqref{p1B} (qualitatively) leads to the
trajectories observed numerically, an example being the curve connecting
$A_0$ with $B_1$ in Fig.~\ref{G2hlconc2}(a). The remaining (and crucial)
question concerns the evolution for $t > t_B$. The numerical simulations show
that the solution $S(t,z)$ continues as an approximate high velocity
transition $\Thi$ for $t > t_B$ (while $R_1$ continues to decrease). In
particular, we find that the entire family of trajectories $S(t,z)$ rejoins
at the final point $C_1$ of the $\Thi$ transition.

Figs.~\ref{G2hlconc2}(a) and~\ref{G2hlconc2}(b) summarize the considerations
and give depictions of a joint low/high velocity spike
transition $\Tco$.
We reemphasize the property that the $t = \mathrm{const}$ curves
are straight lines that coincide with the paths of $\mathcal{T}_{N_1}$ 
curvature transitions.
This `timing' property shows that the spiky structure
that forms under an (approximate) $\Tlo$ is transported by a $R_1$
frame rotation to the spiky structure appearing in (the middle of) an (approximate)
$\Thi$ and eventually `un-forms' according to the $\Thi$ evolution.

\begin{figure}[ht]
\centering 
\includegraphics[height=0.42\textwidth]{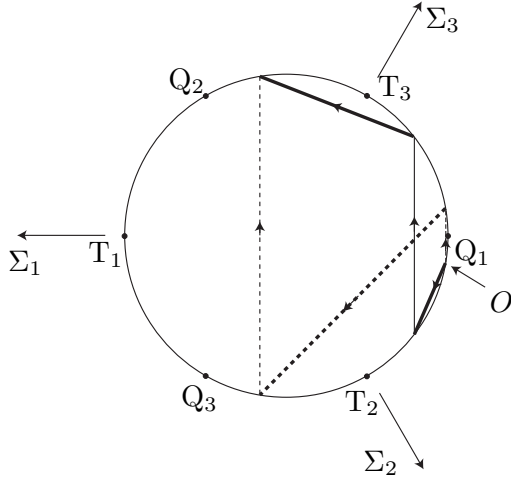}
\caption{A Kasner point $O$ in sector $(132)$ with $1/2 < \ue < 1$
is the initial point of a joint low/high velocity spike transition $\Tco$ (where the continuous
line represents the timeline of the spike surface);
however, there is also a $\mathcal{T}_{R_1}$ frame transition emerging from $O$,
which can be continued with a $\mathcal{T}_{N_1}$ curvature transition and another 
$\mathcal{T}_{R_1}$ frame transition. The two trajectories meet at the same final Kasner point.}
\label{Doublespike}
\end{figure}

A joint low/high velocity spike transition $\Tco$ induces a map
between the initial and the final Kasner state of the transition,
\begin{subequations}
\begin{equation}
\Tco :  \quad 
\ue_- 
\:\:\overset{\Tlo}{\longmapsto} \:\:
\left\{
\begin{array}{c}
1 - \ue_- \\
\ue_- -1
\end{array}\right\}
\:\:\overset{\mathcal{T}_{R_1}}{\longmapsto} \:\: 
\left\{
\begin{array}{cc}
\frac{1}{1 - \ue_-}  & \quad \overset{\Thi}{\longmapsto} \quad  \\
\frac{1}{\ue_- -1} & \overset{\mathcal{T}_{R_3}}{\longmapsto} \; \frac{\ue_-}{1-\ue_-} \:  
\overset{\mathcal{T}_{N_1}}{\longmapsto} 
\end{array}
\right\}
\:\longmapsto\:\:
-\frac{\ue_-}{1 - \ue_-} \:, 
\end{equation}
which, in terms of the standard Kasner parameter coincides with the spike map~\eqref{spikemap}
of $\Thi$ transitions.
The connection between $\Thi$ and $\Tco$ transitions is even stronger: 
Consider Fig.~\ref{Doublespike}. On the one hand, 
the Kasner point $O$ in sector $(132)$ with $1/2 < \ue < 1$
is the initial point of a joint low/high velocity spike transition $\Tco$;
on the other hand, there is a $\mathcal{T}_{R_1}$ frame transition emerging from $O$,
which can be continued with a high velocity transition and another 
$\mathcal{T}_{R_1}$ frame transition; the finite chain \mbox{$\mathcal{T}_{R_1}$--$\Thi$--$\mathcal{T}_{R_1}$}
yields 
\begin{equation}
\ue_- \quad \overset{\mathcal{T}_{R_1}}{\longmapsto} \quad \frac{1}{\ue_-}
\qquad\,\: \overset{\Thi}{\longmapsto} \quad -\frac{1 - \ue_-}{\ue_-}
\quad\,\: \overset{\mathcal{T}_{R_1}}{\longmapsto}\quad -\frac{\ue_-}{1 - \ue_-} = \ue_+\,,
\end{equation}
\end{subequations}
hence the final state of the chain \mbox{$\mathcal{T}_{R_1}$--$\Thi$--$\mathcal{T}_{R_1}$}
coincides with the final state of the joint low/high velocity spike transition $\Tco$.
This corroborates that $\Tco$ and $\Thi$ 
are structures that are on an equal footing (through additional $R_1$ frame rotations).

\begin{figure}[ht]
\centering        
\includegraphics[height=0.44\textwidth]{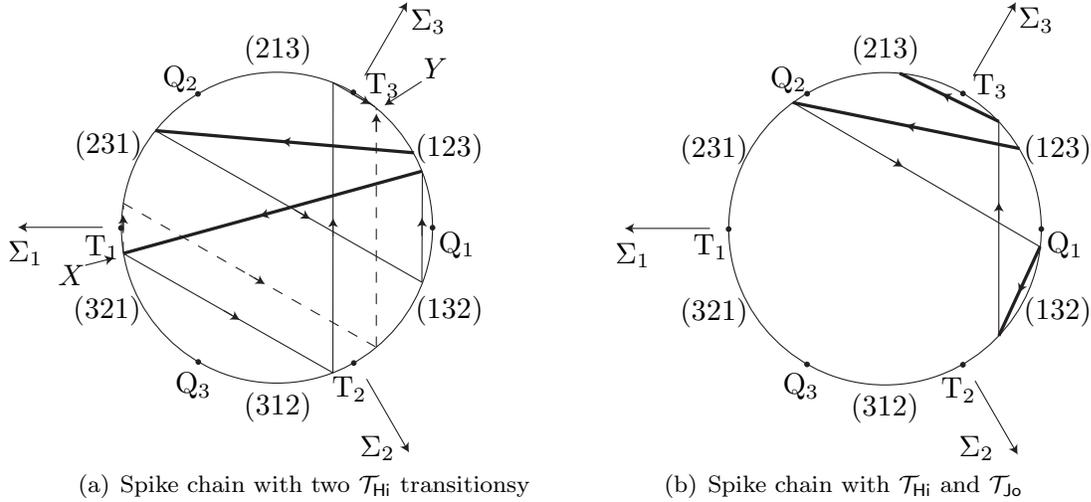}
        \caption{Examples of (parts of) spike chains. In (a) the continuous lines
          represent the spike timeline ($z=0$ trajectory) 
          of $\Thi$--$\mathcal{T}_{R_3}$--$\mathcal{T}_{R_1}$--$\Thi$--$\mathcal{T}_{R_3}$--$\mathcal{T}_{R_1}$--$\mathcal{T}_{R_3}$;
          the dashed lines are $\mathcal{T}_{R_1}$--$\mathcal{T}_{R_3}$--$\mathcal{T}_{R_1}$.
          In (b) we have $\Thi$--$\mathcal{T}_{R_3}$--$\Tco$ (where $\Tco$ consists of three pieces: $\Tlo$--$\mathcal{T}_{R_1}$--$\Thi$).}
           \label{G2hlconc}
\end{figure}

A \textit{spike chain} is an in general infinite concatenation of spike transitions
and frame transitions;
the `chain links' are high velocity spike transitions $\Thi$
and joint low/high velocity spike transitions $\Tco$ and $\mathcal{T}_{R_1}$ and $\mathcal{T}_{R_3}$
frame transitions.
Examples of (short parts of) spike chains are given in Fig.~\ref{G2hlconc}.
The role of spike chains for the dynamics of (generic) $G_2$ models
is suggested by numerical experiments: Asymptotically, solutions $S(t,z)$ 
are \textit{approximate spike chains}, i.e., solutions shadow
spike chains (with an increasing degree of accuracy).
The role of spike chains for generic $G_2$ models is thus
exactly analogous to the role of BKL chains for spatially homogeneous models.

These considerations strongly suggest that the `permanent' asymptotic features 
(`permanent spikes') of $T^3$ Gowdy models, 
which have been emphasized in the literature, are irrelevant for general $G_2$ models
and generic singularities: The permanent features of $T^3$ Gowdy are merely
transient features in the generic context.
Spatial structures form and `un-form' recurrently; however, 
during these `oscillations', the spike widths shrink to zero size.
Oscillations are oscillations between Kasner states; 
the asymptotic dynamics of a spike timeline is represented by 
the spike trajectory ($z= 0$) of a spike chain; 
the asymptotic dynamics of a generic spatial point 
is represented by a BKL chain (which corresponds 
to the trajectory on the `boundary' $|z| = \infty$ of a spike chain).

We conclude with two remarks.
First, consider Fig.~\ref{Doublespike} and recall
that the final state of a low/high velocity spike transition $\Tco$
and the chain \mbox{$\mathcal{T}_{R_1}$--$\Thi$--$\mathcal{T}_{R_1}$}
coincide. It is highly plausible that
`mixed' (or `double') transitions exist, in close analogy with, e.g.,
double frame transitions (involving $R_1$ and $R_3$ at the same
time) and mixed curvature/frame transitions (involving $N_1$ and
frame variables at the same time), see~\cite{heietal09}.
This further stress the close relationship between $\Tco$ and $\Thi$
and identifies $\Tco$ and \mbox{$\mathcal{T}_{R_1}$--$\Thi$--$\mathcal{T}_{R_1}$} as
the limiting cases (boundary) of an entire family of transitions.
However, by analogy with the case of double frame and mixed curvature/frame transitions 
we expect that, asymptotically and generically, merely the single spike
transitions, $\Tco$ and $\Thi$, will occur in a spike chain. 
Second, we note that in the spatially homogeneous cases $\mathrm{VI}_{-1/9}$,
VIII, and~IX, solutions cannot converge to Kasner states (other than the
Taub points---in the LRS case); all solutions are oscillatory.
We expect (almost) the same to be true in the general $G_2$ case:
While in the OT case, solutions converge to generalized Kasner states,
we expect that (almost) all solutions are oscillatory in the general $G_2$ case.

%

\section{Concluding remarks}\label{sec:concl}

In this paper we have discussed the asymptotic dynamics of
$G_2$ models toward a singularity. 
We have given heuristic arguments and strong numerical evidence
showing that the well-known permanent spatial structures (spikes) 
arising in special models such as the $T^3$ Gowdy models 
are absent in general $G_2$ models: 
General $G_2$ models possess oscillatory singularities,
which are characterized by the recurring formation and `un-formation' of spiky features
whose width rapidly shrinks towards the singularity; the asymptotic
dynamics is represented by infinite spike chains 
and thus associated with sequences of Kasner epochs.

The world line of the spike surface is characterized by 
an oscillation between Kasner epochs induced 
by the map that in terms of the 
gauge-invariant Kasner parameter $u$ is given by
\begin{equation}\label{spikemap2}
u \rightarrow
\begin{cases}
u - 2 & u \in [3 ,\infty)\,, \\
(u -2)^{-1} & u \in [2,3] \,,\\
\left((u -1)^{-1} - 1 \right)^{-1} & u \in [ 3/2 ,2] \,,\\
(u -1)^{-1} - 1 & u \in [1, 3/2]\,,
\end{cases}
\end{equation}
which is in contrast to the usual Kasner map
\begin{equation}\label{BKLmap2}
u \rightarrow
\begin{cases}
u - 1 & u \in [2 ,\infty)\,, \\
(u - 1)^{-1} & u \in [1,2] \,.
\end{cases}
\end{equation}
In a subsequent paper~\cite{heiuggtocome1} we will we analyze statistical properties of
the Kasner oscillations of the spike world line 
by investigating the map~\eqref{spikemap2} in detail.

Although we expect that non-local (i.e., non-BKL) dynamics 
will be asymptotically confined to spike surfaces, which is due to the rapidly shrinking spike widths, 
and that these spike surfaces constitute a set of measure zero of all spatial points, 
an understanding of recurring spikes is still crucial for our
understanding of generic spacelike singularities:
\begin{itemize}
\item 
    The key
    solutions that describe BKL and non-BKL behavior are intimately
    related with each other through solution generating algorithms, which
    create a hierarchical order for these solution; this strongly hints at the
    existence of hidden symmetries, which are thus a common theme in our quest to understand
    generic spacelike singularities.\footnote{The solution generating
    techniques used to obtain the spike solutions in a hierarchical
    manner involve the OT line element in an Iwasawa frame,
    i.e.~\eqref{G2ot}, and in the area time gauge. The solutions are
    obtained by alternately applying the so-called Gowdy-to-Ernst (GE)
    transformation and a special frame rotation (FR) (which is obtained by a
    rotation of the symmetry coordinates $x$ and $y$ by $-\pi/2$).
    Starting with the Kasner solution in diagonal form and performing a
    GE transformation yields the Kasner solution in non-trivial OT form;
    a FR yields the frame rotated Kasner solution associated with
    $\mathcal{T}_{R_3}$; subsequent application of a GE transformation
    gives the Bianchi type II vacuum solution in a Fermi frame; 
    applying the FR again yields a `spikily rotating' Bianchi type II vacuum
    solution, known as a false spike solution; acting on this with the GE
    transformation
    results in the explicit spike solution; for details, see~\cite{lim08}.} 
    The case for hidden symmetries is further strengthened by
    the observation that the `spatial boundary' (i.e., $|z| \rightarrow \infty$) 
    of the spike chains are the BKL chains. 
    Since the spike surface trajectories (non-BKL dynamics) meet the trajectories of the other 
    spatial points (BKL dynamics) periodically, there is a close relation
    between the map~\eqref{spikemap2} and the Kasner map~\eqref{BKLmap2}.  
    Accordingly, the BKL scenario seems to be part of
    a greater picture which is needed for a complete understanding of
    generic singularities.
\item Proofs about generic spacelike
    singularities \emph{must} take into account recurring spikes, since
    estimates (e.g., of spatial derivatives) are heavily affected. 
\item Spikes are associated with the zeros of certain functions (e.g., the variable $N_1$).
    Special initial data (like the symmetric data considered in this paper)
    can fix the location of a spike; however, in general, spikes need not
    be present initially but will form when a function goes through zero
    and move in accordance with the evolution of the functions and its zero(s).
    Unfortunately, this does not answer the question what 
    the \emph{physical}
    reasons for spike \emph{formation} are. Our lack of knowledge and
    intuition for spike formation prevents us from even an educated guess about how many
    spikes form generically. At the moment we cannot exclude that
    a dense set of spikes can form toward generic spacelike
    singularities. BKL asymptotic dynamics would still be
    generic, but the BKL scenario would certainly obtain an unexpectedly significant non-BKL counterpart.
\end{itemize}

Let us turn to numerical issues. At present, numerical accuracy regarding
the simulation of spikes can be categorized into three levels:
\begin{itemize}
\item At the crudest level of numerical accuracy there are  not enough
    grid points to simulate spikes correctly; moreover, spikes are
    artificially produced and annihilated.
\item At the intermediate level of numerical accuracy there  are enough
    grid points to accurately describe one spike transition
    correctly, but numerical convergence is not achieved for simulations
    that are supposed to follow a concatenation of spike transitions. 
    The first indication of
    non-convergence is the difference in the timing of transitions. The
    intermediate level yields quantitatively correct results for
    simulations of one spike transition, and qualitatively correct
    results for simulations of short spike chains.
\item Finally, at the highest level of numerical accuracy there are
    enough grid points to obtain numerical convergence for longer
    simulations that cover short spike chains. But even these
    simulations are currently limited to two or three spike transitions.
    We are confident that improvement can come from the heuristic insights gained in this paper.
\end{itemize}

With the benefit of hindsight we are in a position to reproduce and assess what was
actually achieved in the numerical simulations of different papers. Early numerical work on
`spiky features' has to be considered as having crude numerical accuracy,
thus belonging to the first category, but it nevertheless managed to tie
spiky features to analytic conditions in special models such as the $T^3$
Gowdy models, as exemplified in~\cite{bermon93}. Hern~\cite{her99} and Berger
{\it et al} (see~\cite{ber02} and references therein) also studied general
$G_2$ models with $T^3$ topology numerically, but again, even though Hern
used mesh refinement, numerical accuracy was not sufficient to resolve
several spike transitions correctly.\footnote{Hern also simulated spikes in
models with a single spacelike Killing vector~\cite{her99}.} It was not until
2003 that the first qualitatively correct spike simulation was done by
Garfinkle and Weaver~\cite{garwea03} in the case of $T^3$ Gowdy models,
described by means of $P_{,\tau}$ and the speed $v$ (see
Appendix~\ref{app:OT} and Fig.~\ref{hlconc} for a description in terms of
projections on $(\Sigma_1,\Sigma_2,\Sigma_3)$-space).
The first qualitatively correct spike simulation in the general $G_2$ case
was first done in 2005 by Andersson {\it et al}~\cite{andetal05}. This
qualitative picture was confirmed quantitatively and linked to
the Lim's explicit solutions~\cite{lim08} in 2009 by Lim {\it et
al}~\cite{limetal09}. The key idea to simulate several spike
transitions accurately was to use the explicit spike solutions to
design a grid that zoomed in on a single recurring spike. Despite this, only a few
high (but no low) velocity spike transitions were followed.

The work~\cite{limetal09} makes it clear what a formidable numerical
challenge recurring spikes pose. It is important to note that all quantitatively correct
spike simulations so far involve recurring spikes that are forced by
special initial conditions to be fixed in space (non-moving spikes).
The spike surface thus obtains a fixed value of the spatial coordinate, e.g.,
$z_{\mathrm{Spike}} = 0$. In this paper we have pushed the current numerical
frontier in the context of the recurrence of `\emph{non-moving spikes}' by numerically correct
simulations of a chain of spike transitions that also involve low velocity
spike solutions. The main challenge in this context is that low velocity spike solutions 
require numerical accuracy at super horizon scales.

In general, recurring spikes will not be spatially fixed, instead they will
move, i.e., $z_{\mathrm{Spike}} = z_{\mathrm{Spike}}(t)$. These
`\emph{moving spikes}' are not directly described by the explicit spike solutions~\eqref{metricspike}.
However, it is possible that moving spikes
`\emph{asymptotically freeze}' in space which implies that the description
of the recurrence of these spikes in terms of spike chains straightforwardly 
applies.\footnote{There exists heuristic evidence
for asymptotically frozen spikes. In the
billiard picture, a `dominant wall' that has a hole in it at an isolated value of a
spatial coordinate corresponds to a spatially frozen
spike~\cite{dametal03}.} 
The question of asymptotic freezing is closely related to another unexplored area:
Present
numerical techniques are not sufficient to accurately describe the creation 
and the possible annihilation of spikes. This makes it difficult to 
guess whether the number of spikes that form remains finite, or whether,
in general, an infinite number of spikes, possibly a dense set,
is created in the asymptotic regime. It is doubtful whether
a sound numerical investigation of this question is possible
unless the numerics is supported by analytical insights.

There are many examples in general relativity that show that one must go
beyond a given context in order to understand it, especially as regards
asymptotics. 
This is particularly relevant as regards topology.
For instance, Bianchi type~I and~II models 
are essential building blocks for the understanding of the singularities
of the more general Bianchi models; it is important to note that 
this is irrespective of the fact that
these Bianchi models have spatial topologies that are completely different
from those of Bianchi types I and II. 
Yet another example is
provided by the present paper: Understanding the singularities of
$G_2$ models requires to go beyond $T^3$ topology, since
the essential building blocks, the explicit high and low velocity spike solutions
do not admit such a topology. Special models like these provide \textit{local} 
building blocks (where local may refer to the particle horizon scale) 
for the asymptotic description of more general models, or even generic ones. 

The relationship between local and global issues deserves special attention.
It is important to realize that the primary importance of the $G_2$ models as
regards generic singularities is not the models themselves, but the crucial property 
that the $G_2$ equations are those of an invariant boundary subset of
the general conformally Hubble-normalized 
state space (associated with the Einstein equations without symmetries),
which we call the \emph{partially local $G_2$ boundary}.\footnote{The partially local $G_2$ boundary, previously
referred to as a the partially silent boundary subset, yields, e.g., spiky behavior
that is still associated with asymptotic silence, but also solutions with
singularities that break asymptotic silence, which are characterized 
by the existence of directions in which particle horizons extend to infinity,
see~\cite{limetal06}.} This subset is obtained by setting certain frame variables to
zero to achieve $\parb_1(*)=\parb_2(*)=0$ in the field equations, where
$\parb_1$ and $\parb_2$ are Hubble-normalized spatial frame derivatives; by this, 
the equation induced on the partially local $G_2$ boundary are the same as those for the $G_2$ models,
where, however, the constants of integration may now depend on the spatial coordinates
$x$ and $y$. The situations is thus completely analogous to the relationship
between the local boundary and the spatially homogeneous models. 
Since what survives of the $G_2$ models in the context of the 
general conformally Hubble-normalized state space
is merely the equations (on a boundary subset), 
any global considerations, which may be of interest for the $G_2$ models
themselves, are probably not particularly relevant in the general context (for general models
there typically are other global considerations that are relevant). As
regards singularities it is the \emph{local} results, concerning domains of dependence 
in the vicinity of the singularity, 
that are of importance for, e.g., cosmic censorship in the generic case for models without symmetries.
For further discussion on topological issues in the case of $G_2$ models, see
Appendix~\ref{app:topology}. 

Going beyond the $G_2$ assumption and looking at models with fewer isometries
does not only shift the attention to the partially local $G_2$ boundary, it
leads to completely new challenges as well. Spike surfaces, within the
context of the partially local $G_2$ boundary, are no longer planes, and they can intersect
in curves that lead to different spike dynamics. Similarly, if one has no
symmetries at all, spike curves may intersect at points, which again may lead to new
spike dynamics. At present it is not known whether such intersections
persist or recur, or if they are transient and hence irrelevant asymptotically.%
\footnote{Weak numerical evidence suggests that
intersections only occur momentarily, and hence that the BKL picture in
combination with $G_2$ spike oscillations might
capture all essential features of generic spacelike singularities.}
The above issues illustrate again that spike dynamics poses a 
formidable numerical and analytical challenge. It is clear that we are only 
beginning to understand the rich structure of generic singularities, and
the underlying physical reasons for their existence and characteristics.

\subsection*{Acknowledgments}
We gratefully acknowledge the hospitality of the Erwin Schr\"odinger Institute, 
where part of this work was done. 
CU is supported by the Swedish Research Council project number 621-2009-4163.

\begin{appendix}

\section{Appendices} 
\label{app:methubfield}

\subsection{$\bm{G_2}$ models and topology}\label{app:topology}

A $G_2$ model is a spacetime admitting
a two-dimensional Abelian isometry group.
In the literature the spacetime is often assumed to be globally hyperbolic and
the action to be effective on spacetime Cauchy hypersurfaces with 
two-dimensional principal orbits.
A substantial part of the rigorous results concerns $G_2$ models with a 
compact Lie group which is identified with \mbox{$T^2 = U(1) \times U(1)$}; for this
reason these models are occasionally referred to as `$T^2$-symmetric spacetimes'. By
assuming compactness of the Cauchy hypersurfaces  it follows that the
topology has to be $T^3$, $S^2 \times S^1$, $S^3$, or that of one of the Lens
spaces $L(p;q)$ (which are quotient spaces of $S^3$), see~\cite{chr90}. In
the cases of $S^2 \times S^1$, $S^3$, and $L(p;q)$, the group action must
have fixed points (which correspond to zeros of the Killing vector fields),
and hence the two `twist constants'
\begin{equation}
c_A = \mathrm{vol}_{a b c d} X_1^a X_2^b \nabla^c X_A^d \qquad (A=1,2)
\end{equation}
of the two commuting Killing vector fields $X_1$ and $X_2$ vanish. (The twist
constants are spacetime constants, i.e., $d\, c_A = 0$, if the Einstein
vacuum equations are imposed~\cite{ger72,chr90}.) In other words, in these
cases the action is necessarily `\textit{orthogonally transitive}'
(OT)~\cite{waiell97}, i.e., the 2-spaces orthogonal to the group orbits are
surface forming.

The spatially compact topology that is of particular interest in the context
of generic singularities is $T^3 = S^1 \times T^2$ (and its covering
$\mathbb{R} \times T^2$), because in this case the twist constants do not
vanish in general, i.e., in the $T^3$ case, $c_1 = c_2 = 0$ is a restriction.
The (standard) $T^3$ case arises from the $\mathbb{R} \times T^2$ case by
identifying $\{0\}\times T^2$ and $\{2\pi\} \times T^2$ by means of periodic
boundary conditions for the metric and the extrinsic curvature. (The
corresponding OT models are the $T^3$ Gowdy models~\cite{gow74}.) It is also
possible to consider non-trivial torus bundles over $S^1$, which correspond
to identifying $\{0\}\times T^2$ and $\{2\pi\} \times T^2$ through
non-trivial $SL(2,\mathbb{Z})$ transformations. In this case, the action of
$T^2$ is global on the covering space $\mathbb{R} \times T^2$ but merely
local on $T^3$ (i.e., there do not exist global Killing vector fields), see,
e.g.,~\cite{ren12},\footnote{These considerations are also relevant in
connection with the specialization of $G_2$ models to spatially homogeneous
Bianchi models~\cite{ren12}.} and the metric and the extrinsic curvature
satisfy non-trivial boundary conditions.

It is possible to w.l.o.g.\ assume the vanishing of one of the twist
constants; namely, if $c_1 \neq 0$ and $c_2 \neq 0$, there are linear
combinations $X_1'$ and $X_2'$ of $X_1$ and $X_2$, which are again Killing
vector fields, such that the associated twist constants are $c_x = 0$ and
$c_y \neq 0$. We introduce local coordinates\footnote{Unless $X_1$, $X_2$ and
$X_1'$, $X_2'$ are related by an $SL(2,\mathbb{Z})$ transformation, $x$ and
$y$ are not standard coordinates on $T^2$; however, this does not affect the
equations; see~\cite{chr90}.} such that $X_1' \equiv \partial_x$ and
$X_2'\equiv \partial_y$. Then the metric takes the form~\eqref{metricbarn},
which coincides with the form in~\cite{beretal01, beretal97} by setting $U =
-b^1$, $\log R = -b^1 -b^2$, $\nu = -b^1 - b^3$, $\alpha = N^2 \exp(2 b^3)$;
the shift is set to zero, see also~\cite{andetal04a, ren08}; finally, $A =
-\bar{n}_1$, $G_1 = -\bar{n}_2$, $G_2 = -\bar{n}_3$. In the spatially compact
case, these functions are periodic in $z$ (or, in the case of a non-trivial
torus bundle, satisfy boundary conditions derived from the particular
$SL(2,\mathbb{Z})$ identification). In terms of the metric functions,
cf.~\eqref{metricn}, and the area density $R$, the twist constants are
represented by
\begin{subequations}\label{twico}
\begin{align}
c_x &= 0\:, 
\\
c_y &= R^3 e^{2 b^1 + b^3} N^{-1} \partial_{x^0} n_3 =
e^{-b^1 -3 b^2 + b^3} N^{-1}\partial_{x^0} n_3 \,.
\end{align}
The vanishing of the first twist constant is equivalent to the identity
\begin{equation}\label{nident}
-\partial_{x^0} \bar{n}_2 + \bar{n}_1 \partial_{x^0} \bar{n}_3 =
\partial_{x^0} n_2 - n_3 \partial_{x_0} n_1 = 0 \:,
\end{equation}
\end{subequations}
i.e., $R_2=0$, see eq.~\eqref{R2fieldeq} below. In area time
gauge~\eqref{areatime} the metric takes the form~\eqref{metricarea}, where
$\bar{\alpha}$ satisfies the equation
\begin{equation}\label{alphabar}
\partial_t \bar{\alpha} = e^{2 b^2 - 2 b^3} \bar{\alpha}^2 c_y^2 \:,
\end{equation}
see~\cite{beretal97}, and $c_y$ becomes $c_y = -e^{-2 b^2 + 2 b^3}
\bar{\alpha}^{-1/2} \partial_t n_3$.

The OT case is characterized by vanishing twist constants;
through~\eqref{twico} this leads to functions 
$n_2$ and $n_3$ that are independent of time. Using the coordinate freedom $x
\mapsto x + f_1(z)$, $y \mapsto y + f_2(z)$ we are able to
set 
$n_2 = n_3 \equiv 0$ without loss of generality. Likewise, we obtain that
$\bar{\alpha}$ is independent of time from~\eqref{alphabar}. Using the
remaining coordinate freedom $z \mapsto f_3(z)$ and redefining $b^3$ we may
thus set $\bar{\alpha} \equiv 1$, which corresponds to a lapse
function~\eqref{lapse}. Note that setting $\bar{\alpha} \equiv 1$ is
impossible in the general case (with $c_y \neq 0$) as long as the area time
gauge is enforced; on the other hand, abandoning~\eqref{areatime} makes the
`conformal gauge' (or: `null cone gauge'~\cite{elsetal02}) $\bar{\alpha}
\equiv 1$ possible~\cite{beretal97}.


\subsection{Metric variables and Hubble-normalized variables in the $\bm{G_2}$ case}\label{app:methub}

In the $G_2$ case, the general relations of Appendix A of~\cite{heietal09}
reduce to the following relationship between the metric variables $b^1$,
$b^2$, $b^3$, $n_1$, $n_2$, $n_3$, and the Hubble variable $H$ and the
conformally Hubble-normalized state space variables:
\begin{subequations}\label{relations}
\begin{alignat}{2}
\label{e3n1}
E_3 &= H^{-1}\exp(b^3)\,, & \qquad
H & = -\textfrac13 N^{-1} \partial_{x^0}(b^1 + b^2 + b^3)\,,\\
r &= - H^{-2}\exp(b^3)\partial_{z}H\,, & \qquad
\dot{U} & = -r + (N H)^{-1}\exp(b^3)\partial_{z}N\,, \\
\label{Arq}
A  &= r + \textfrac12 H^{-1}\exp(b^3)\partial_{z}(b^1 + b^2)\,,  & \qquad
q & = - 1 - N^{-1}H^{-2}\partial_{x^0} H\,.
\end{alignat}
For simplicity we have set $E_3 = E_3^{\:3}$, $A=A_3$, $r=r_3$, and
$\dot{U}=\dot{U}_3$. These quantities refer to an Iwasawa frame; functions
depend on $x^0$ and $z$ alone. Furthermore,
\begin{align}
\label{saeq}
\Sigma_\alpha &= - 1 - (N H)^{-1}\partial_{x^0}(b^\alpha)  \qquad (\alpha =1,2,3)\,,\\
R_1 &= -\Sigma_{23} = -\textfrac12 (N H)^{-1}\exp(b^3-b^2)\partial_{x^0}(n_3)\,,  \\
R_2 & = \Sigma_{31} = \textfrac12 (N H)^{-1}\exp(b^3-b^1)
\left[\partial_{x^0}(n_2) - n_3\partial_{x^0}(n_1)\right] \,, \label{R2fieldeq}\\
R_3 & =-\Sigma_{12}  =-\textfrac12 (N
H)^{-1}\exp(b^2-b^1)\partial_{x^0}(n_1)\,,  \\
N_1 &= H^{-1}\exp(b^2 + b^3 - b^1)\partial_{z}(n_1)\,,  \qquad\quad
N_{12} = \sfrac12 H^{-1}\exp(b^3)\partial_{z}(b^1 -b^2)\,.
\end{align}
\end{subequations}
Adapting the spatial coordinates to achieve the vanishing of the first twist
constant leads to $\partial_{x^0}n_2 - n_3\partial_{x^0}n_1 = 0$,
see~\eqref{nident}, and thus $R_2  = 0$. In area time gauge, the sum $b^1 +
b^2$ does not depend on $z$, hence $A \equiv r$ in that case. The lapse $N$
is unspecified in general; for OT models in area time gauge, however, we have
$N = -\exp(-b^1 - b^2 - b^3)$, see~\eqref{lapse}.

\subsection{Vacuum field equations in the $\bm{G_2}$ case}\label{app:fieldeqG2}

The vacuum field equations for the conformally Hubble-normalized variables in
the $G_2$ case are conveniently divided into decoupled equations and a
coupled system of evolution equations and constraints.

The \emph{decoupled equations} (for the metric variables and the Hubble scalar) read
\begin{subequations}
\begin{xalignat}{2}
& \parb_0 b^\alpha = - (1 + \Sigma_\alpha)\,,\quad (\alpha =1,2,3)
& & \parb_0 n_1 = -2R_{3}\exp(b^2 - b^1) \:,\\
& \parb_0 n_2 = n_3\parb_0n_1\,,& \quad & \parb_0 n_3 = -2R_{1}\exp(b^3 - b^2)\,, \\
& \parb_0 H = -(1+q)H\,,  \\[1ex]
& \parb_3(b^1 - b^2) = 2N_{12}\,, & \quad
& \parb_3(b^1 + b^2) = 2(A - r)\,, \\
& \parb_3n_1 = N_1\exp(b^1 - b^2)\,,&\quad
& \parb_3N = (\dot{U} + r)N\,, \\
& \parb_3H = -rH\,, &&
\end{xalignat}
\end{subequations}
where the conformal frame vectors are
\begin{equation}
\parb_0 \equiv (N H)^{-1}\partial_{x^0}\,,\qquad \parb_3 \equiv E_3 \partial_{z}\,.
\end{equation}

The \emph{coupled evolution equations} for the conformally Hubble-normalized variables are
\begin{subequations}\label{devoleq}
\begin{align}
\label{parb0E3}
\parb_0 E_3 &= (q - \Sigma_{3})E_3,\\
\parb_0\Sigma_{1} &= -(2 - q)\Sigma_1 + 2R_3^2 - {}^3\!{\cal S}_1
- \sfrac{1}{3}(\parb_{3} + A + 3N_{12})(\Udot + 2r) -\sfrac13(\Udot^2 - 2r^2)\, ,\\
\parb_0\Sigma_{2} &= -(2 - q)\Sigma_2 - 2R_{3}^2 + 2R_{1}^2 - {}^3\!{\cal S}_2
- \sfrac{1}{3}(\parb_{3} + A - 3N_{12})(\Udot + 2r) -\sfrac13(\Udot^2 - 2r^2)\, ,\\
\label{Sig3eq}
\parb_0\Sigma_{3} &= -(2 - q)\Sigma_3 - 2R_{1}^2 - {}^3\!{\cal S}_3
+ \sfrac{2}{3}(\parb_{3} + A)(\Udot + 2r) + \sfrac23(\Udot^2 - 2r^2)\, ,\\
\parb_0R_{1} &= -(2 - q + \Sigma_2 - \Sigma_3)R_{1}\, ,\\
\label{R3eq}
\parb_0R_{3} &= -(2 - q + \Sigma_1 - \Sigma_2)R_{3} + {}^3\!{\cal S}_{12}
- \sfrac12N_1\,(\Udot + 2r)\, ,\\
\parb_0 A &= (q - \Sigma_3)A +
\textfrac{1}{2}(\parb_3 + \Udot)(2q + \Sigma_3),\\
\parb_0 N_1 &=  (q + 2\Sigma_1)N_1 -
2(\parb_3 + \Udot + 2N_{12})R_3\, ,\\
\parb_0 N_{12} &=  (q - \Sigma_3)N_{12} -
\sfrac12(\parb_3 + \Udot)( \Sigma_1 - \Sigma_2)\, .
\end{align}
\end{subequations}

The \textit{coupled constraint equations} take the form
\begin{subequations}\label{dconstreqG2}
\begin{align}
0 &= 1 - \Sigma^2 - \Omega_k - \sfrac13(2\parb_3 - 4A + r)\,r\, ,\label{dGauss2}\\
0 &= (\parb_3 -3A + 2r + N_{12})R_{1}\, ,\label{dCodazzi2}\\
0 &= 2r + (\parb_3 -3A + 2r)\Sigma_3 + (\Sigma_1-\Sigma_2)N_{12}
+ N_1R_{3}\,.\label{dCodazzi3}
\end{align}
\end{subequations}

In these equations we have used the abbreviations
\begin{subequations}\label{threecurv2}
\begin{align}
\Omega_k &= \sfrac{1}{12}(N_1^2 + 4N_{12}^2) - \sfrac13(2\parb_3 - 3A)A\, ,\\
{}^3\!{\cal S}_1 &= \sfrac23(N_1^2 + N_{12}^2) - 2N_{12}\,A - \sfrac13\parb_3(A - 3N_{12})\,,\\
{}^3\!{\cal S}_2 &= -\sfrac13(N_1^2 - 2N_{12}^2) + 2N_{12}\,A - \sfrac13\parb_3(A + 3N_{12})\,,\\
{}^3\!{\cal S}_3 &= -\sfrac13(N_1^2 + 4N_{12}^2) + \sfrac23\parb_3\,A\,,\\
{}^3\!{\cal S}_{12} &= -\sfrac12(\parb_3 - 2N_{12} - 2A)N_1\,,\\
q &= 2\Sigma^{2} - \textfrac{1}{3}
\left[\parb_3 + \Udot -2 (A -r)\right](\Udot + r)\, ,\label{q}\\
\Sigma^2 &= \sfrac16\!\left(\Sigma_1^2 + \Sigma_2^2 + \Sigma_3^2 +
2R_{1}^2 + 2R_{3}^2\right)\,.
\end{align}
\end{subequations}

The conformally Hubble-normalized equations are not the most useful equations
in the $G_2$ context. Instead one can adapt to the special structure of these
models and use, e.g., conformally area-expansion-normalized variables,
see~\cite{elsetal02,lim04}. However, since we are interested in how the $G_2$
case fits into the general context without symmetries we have focused on the
conformal Hubble-normalized approach in this paper.

\subsection{OT models: Asymptotic velocity dominance}\label{app:OT}

In this section we give a brief overview of existing rigorous results for
$T^3$ Gowdy vacuum models. Since these results are intimately connected with
the concept of (asymptotic) velocity, we begin by introducing this particular
diagnostic tool. The metric~\eqref{G2ot} in area time gauge is sometimes written
in the form
\begin{equation}\label{GowdyT3}
d s^2 = -e^{(t - \lambda)/2}\,\left( {-e^{-2 t}} dt^2 + d z^2 \right)
+ e^{P-t} \left( d x + Q d y\right)^2 + e^{-P-t} d y^2 \:,
\end{equation}
see, e.g., equation~(15) in~\cite{rin10}, where $(t,x,y,z)$ corresponds to
$(\tau,\sigma,\delta,\theta)$. Obviously,~\eqref{GowdyT3} is of the
form~\eqref{G2ot} with $2 b^1 = {-P} + t$, $2 b^2 = P + t$, $2 b^3 =
\lambda/2 - t/2$, $n_1 = Q$. There exists a number of diagnostic tools
connected with~\eqref{GowdyT3}, which were originally developed for $T^3$
Gowdy models but are useful for all OT models in area time gauge. In
particular one can define kinetic and potential energy densities according
to~\cite{rin06}
\begin{equation}\label{kinpot}
\mathcal{K} = \mathcal{K}(t,z) = (P_{, t})^2 + (e^{P} Q_{, t})^2 \,,
\qquad
\mathcal{P} = \mathcal{P}(t,z) = (e^{-t} P_{, z})^2 + (e^{P-t} Q_{, z})^2 \,,
\end{equation}
and a \textit{velocity} $v$ as the square root of $\mathcal{K}$; more
appropriately, the velocity should be referred to as a speed (since it is
non-negative by definition). The quantities $\mathcal{K}, \mathcal{P}$, and
$v$ are intimately connected with the Hubble-normalized variables.
In~\cite{andetal04} it was noted that the velocity $v$ is
\begin{equation}\label{veloc}
v = v(t,z) =  \sqrt{\mathcal{K}}  =
\frac{\sqrt{\frac14(\Sigma_1 - \Sigma_ 2)^2 + R_3^2}}{1 - \frac12\Sigma_3}\,
= \sqrt{3}\frac{\sqrt{\Sigma^2 - \frac14\Sigma_3^2}}{1 - \frac12\Sigma_3}\,.
\end{equation}
At the same time we obtain
\begin{subequations}\label{hubbinkinpot}
\begin{alignat}{3}
\pe_1 & = \frac{2 (1 - P_{, t})}{3 + \mathcal{K} + \mathcal{P}}
\,,\qquad & \pe_2 & = \frac{2 (1 + P_{, t})}{3 + \mathcal{K} +
\mathcal{P}} \,,\qquad &
\pe_3 & = \frac{-1 + \mathcal{K} + \mathcal{P}}{3 + \mathcal{K} + \mathcal{P}} \,,\\[0.4ex]
R_3 & = 6  \,\frac{e^{P} Q_{, t}}{3 + \mathcal{K} +
\mathcal{P}} \,,\qquad & N_1 & = 12 \,\frac{e^{P-t} Q_{, z}}{3
+ \mathcal{K} + \mathcal{P}} \,,\qquad & N_{1 2} & = {-6} \,
\frac{e^{-t} P_{, z}}{3 + \mathcal{K} + \mathcal{P}} \,,
\end{alignat}
from which we infer that
\begin{equation}
\pe_1^2 + \pe_2^2 + \pe_3^2 =
1- 8\, \frac{e^{2 P} Q_{, t}^2 + \mathcal{P}}{(3 + \mathcal{K} + \mathcal{P})^2}\,,
\qquad
\Sigma^2 = 1 - 12\, \frac{\mathcal{P}}{(3 + \mathcal{K} + \mathcal{P})^2} \,.
\end{equation}
\end{subequations}

A considerable part of the analytic rigorous work on spacelike singularities
has been influenced by the work by Eardley {\it et al}~\cite{earetal72},
which has resulted in a focus on \emph{asymptotic velocities}. The term
asymptotic velocity refers to the limit
\begin{equation}\label{velocinf}
v_\infty = v_\infty(z) = \lim_{t\rightarrow \infty} v(t,z) =
\lim_{t\rightarrow \infty} \sqrt{\mathcal{K}(t,z)} = |2\ue + 1|\:.
\end{equation}
In the context of the $T^3$ Gowdy models, the existence of this limit has
been proved (but we emphasize that this limit does in general not exist for
$G_2$ models without the OT assumption, see section~\ref{sec:genG2}).
Specifically, for each smooth solution $(P,Q)$, $v_\infty$ is an upper
semicontinuous function of $z$~\cite{rin06}. Furthermore, $e^{2 P} Q_{, t}^2$
and $\mathcal{P}$ converge to zero as $t\rightarrow \infty$ and $P_{, t}^2
\rightarrow v_\infty^2$ ($t\rightarrow \infty$). (This holds not merely
pointwise in $z$ but uniformly on the shrinking particle horizons $[z -
e^{-t}, z + e^t]$ associated with a $z = \mathrm{const}$ timeline; we refer
to our discussion of particle horizons in section~\ref{sec:OT}. Evidently,
the spatial topology is irrelevant in this context.) Following Moncrief and
coworkers, the asymptotic velocity $v_\infty$ has been used extensively as a
diagnostic tool see,
e.g.,~\cite{grumon93,kicren98,rin10,rin06,iseetal90,chachr04}. To resolve the
ambiguity in the limit of $P_{,t}$, following~\cite{renwea01,kicren98} one
may define the asymptotic velocity component
\begin{equation}
k = \lim_{t\rightarrow \infty} P_{, t} = \lim_{t\rightarrow \infty} \frac{\Sigma_1 - \Sigma_ 2}{2(1 -
\frac12\Sigma_3)} = 2\ue + 1\,,
\end{equation}
i.e., $|k| = v_\infty$; it follows that $0 < k < 1$ corresponds
to $-\frac12 < \ue < 0$, i.e., to the stable sector (312).

The convergence statements are commonly subsumed under the term `asymptotic
velocity dominance', and the existence of the limit~\eqref{velocinf} is
referred to as convergence of solutions to an asymptotically velocity
dominated state. As a consequence of~\eqref{hubbinkinpot}, however, the
convergence statements simply mean that each solution converges to a
generalized Kasner metric with a semi-continuous Kasner parameter $\ue =
\ue(z)$.

The convergence of solutions to the Kasner circle $\mathrm{K}^\ocircle$ is
understood in quite some detail in the $T^3$ Gowdy case. Building
on~\cite{grumon93} and the results of Rendall and
coworkers~\cite{kicren98,ren00}, Ringstr\"om proved that solutions $(P,Q)$
with $0 < k < 1$ (which implies $k = v_\infty$) possess asymptotic expansions
of the form
\begin{equation}\label{asyexpans}
P(t,z) = v_{\infty}(z) t + \phi(z) + u(t,z) \,,
\qquad
Q(t,z) = q(z) + e^{-2 v_\infty(z) t} \big( \psi(z) + w(t,z) \big) \,,
\end{equation}
where $u$ and $w$ and its derivatives converge to zero exponentially, see
Prop.~1.5 in~\cite{rin06} for the precise statement.\footnote{The assumption
$0 < k(z_0) < 1$ for some $z_0$ is sufficient to obtain smoothness of
$v_\infty$ and uniform expansions of the type~\eqref{asyexpans} in a
neighborhood of $z_0$.} Note that the expansion~\eqref{asyexpans} represents
the convergence of solutions to the stable sector of the Kasner circle.
In~\cite{chachr04, rin04a,rin04b} criteria on initial data have been derived
that guarantee the assumption $0 < k < 1$ on the asymptotic velocity. By
these theorems, the heuristic reasoning of section~\ref{spikeorbits} that
solutions that are sufficiently close to a generalized Kasner metric on the
stable sector converge to the stable sector is made rigorous in the context
of $T^3$ Gowdy models.

Besides~\eqref{asyexpans} there exist solutions with asymptotic expansions of
the type~\eqref{asyexpans} where, however, the asymptotic velocity $v_\infty$
is discontinuous. These are the solutions that exhibit a `true'
\textit{spike} in the nomenclature of~\cite{rin10}.\footnote{In this context,
the term `true' refers to the non-uniform convergence of scalar functions,
e.g., the Hubble-normalized Kretschmann scalar, to a discontinuous limit.
`False' spikes (which we refrain from discussing) possess a discontinuous
$k$, but $v_\infty$ and curvature scalars are continuous.} Alternatively,
these solutions are said to possess a `permanent' spike. The explicit low
velocity spike solutions~\eqref{spikemaploe} fall into this category, while
the explicit high velocity spike solution~\eqref{spikemape} possess a
constant (and thus continuous) asymptotic velocity. However, we strongly
emphasize that the `permanence' of low velocity spike solutions, i.e.,
convergence (to a discontinuous limit), is true \textit{in the OT context
only}. As seen in section~\ref{sec:genG2}, spikes are not permanent but
transient and recurring features in the general $G_2$ case.

We note that the usefulness of the (asymptotic) velocity in the
characterization of the behavior of OT models is indisputable since it
captures essential aspects of the dynamics of solutions; however, it should
not come as a surprise that $v$ fails to capture \textit{all} aspects of the
dynamics. Furthermore, the role of the velocity is diminished when we go
beyond the OT case where the asymptotic velocity $v_\infty$ does not even
exist since $v$ does not converge, see section~\ref{sec:genG2}. However, we
obtain a clearer picture when we do not restrict ourselves to one particular
degree of freedom but instead consider projections of the Hubble-normalized
state vector onto the 2-dimensional space spanned by
$(\Sigma_1,\Sigma_2,\Sigma_3)$, as is done, e.g., in the present paper.

Finally, the issue of `high' and `low' velocities deserves a comment. In the
Gowdy literature the question of whether a velocity is high or low usually
refers to the asymptotic velocity $v_\infty$. However, if the
limit~\eqref{velocinf} does not exist, as for the general $G_2$ models, this
classification becomes meaningless. In addition, the explicit spike
solutions~\eqref{metricspike} do not easily fall into the `high/low'
asymptotic velocity scheme: For the explicit high velocity solutions the
limit $v_\infty$ is an arbitrary positive number; for the explicit low
velocity solutions the limit is in the interval $(1,2)$ for the spike
timelines and in the interval $(0,1)$ for the $z\neq 0$ timelines. In the
context of the explicit spike solutions it is thus preferable to classify
solutions in terms of their limit $t\rightarrow -\infty$; the corresponding
(anti)asymptotic velocity (as $t\rightarrow -\infty$) is in the interval
$(2,3)$ for the explicit low velocity spike solutions and in the interval
$(3,\infty)$ for the explicit high velocity spike solutions, thus motivating
the present nomenclature for the spike solutions. With hindsight, and with
issues concerning generic singularities and cosmic censorship as ultimate
goals, we find that there is room for improvement concerning the terminology
that is used in much of the literature.

\subsection{Evolution of spike widths}
\label{spikewidthevol}

In this section of the appendix we show the result given in
equation~\eqref{spikeevol}. Consider equation~\eqref{R3eq}. In area time
gauge, since~\eqref{Arq} implies that $A \equiv r$, we have
\begin{equation}\label{R3der}
\mathcal{N}^{-1} \partial_t R_3 = -(2 -q + \Sigma_1 -\Sigma_2) R_3 -
\textfrac{1}{2} \, \parb_3 N_1 + N_1 N_{1 2} - \textfrac{1}{2} N_1 \dot{U} \:,
\end{equation}
where $\dot{U} = \parb_3 \log (-\mathcal{N})$ and where the Hubble-normalized
lapse $\mathcal{N}$ is given by
\begin{equation}
\mathcal{N} = N H = -\frac{1}{2 - \Sigma_3}\:,
\end{equation}
which follows from $b^1+b^2=t$ and~\eqref{saeq} by using that $\Sigma_3 =
-\Sigma_1 - \Sigma_2$. Making use of~\eqref{q} and the
constraint~\eqref{dGauss2} equation~\eqref{Sig3eq} turns into
\begin{equation}
\mathcal{N}^{-1} \partial_t \Sigma_3 = (2-q) (2 - \Sigma_3) -2 R_1^2 \:.
\end{equation}
Therefore, in the OT case, where $R_1 \equiv 0$, we find that
\begin{equation}\label{mathcalNder}
\mathcal{N}^{-1} \partial_t \mathcal{N} = (2-q) \mathcal{N}\:.
\end{equation}
Accordingly, equation~\eqref{R3der} yields
\begin{equation}\label{NR3}
\mathcal{N}^{-2} \partial_t \big(\mathcal{N} R_3 \big) =
-(\Sigma_1 -\Sigma_2) R_3 - \textfrac{1}{2} \, E_3 \,
\partial_z N_1 + N_1 N_{1 2} - \textfrac{1}{2}  N_1 E_3 \,\partial_z \log (-\mathcal{N})  \:.
\end{equation}
Furthermore, from~\eqref{parb0E3} and~\eqref{mathcalNder} we obtain
\begin{equation}
\partial_t (\mathcal{N} E_3) = - \mathcal{N} E_3 \:,
\end{equation}
which entails that $\mathcal{N} E_3 = {-k} e^{-t}$ for some $k = k(z)$; we
set $\kappa = k(0)$.

Consider an OT solution $\mathcal{S}$ that is an approximate OT spike chain
with spike surface $z = 0$ and symmetric functions, i.e., $\Sigma_1$,
$\Sigma_2$, $\Sigma_3$, and $R_3$ are even and $N_1$, $N_{1 2}$ are odd. We
evaluate~\eqref{NR3} at the spike surface $z = 0$ to obtain
\begin{equation}
\partial_z N_1  \,\big|_{z = 0} = -2 (\mathcal{N} E_3)^{-1}
\big|_{z=0}\:\big( \mathcal{N}^{-1} \partial_t
\big(\mathcal{N} R_3 \big) + (\Sigma_1 -\Sigma_2)\, \mathcal{N} R_3 \big) \,\big|_{z=0} \:.
\end{equation}
Insertion into~\eqref{widthdef} yields an expression for the spike coordinate
width $c_z$ according to
\begin{equation}\label{czexpr}
c_z = (\mathcal{N} E_3)^{-1}\,\big|_{z = 0} \,\times\, \mathrm{function}
\big(\text{spike surface orbit $z= 0$}\big) \:.
\end{equation}
Note that the spike surface orbits at $z = 0$ are independent of the width of
the (approximate) spike transition. For an explicit spike solution with width
$\check{c}$ and time offset ${-\check{t}}$, i.e.,~\eqref{metricspike} where
$z$ is replaced by $\check{c} z$ and $t$ by $t - \check{t}$, we have
$\mathcal{N} E_3 = -\check{c}^{-1} e^{-(t-\check{t})}$, which follows
from~\eqref{e3n1} and~\eqref{coordtransf}, while the function of the spike
surface orbits yields ${-e^{-(t-\check{t})}}$.

Suppose that the approximate OT spike chain is approximated in some time
interval by an exact spike transition with parameter $\ue_0$, (inverse) width
$c_0$, and time offset ${-t_0}$; hence, by~\eqref{czexpr},
\begin{equation}\label{c0}
c_0 = -\kappa^{-1} e^t \,e^{-(t-t_0)} \:.
\end{equation}
At a (much) later time, when the solution is approximated by the subsequent
$\Thi$ of the OT chain; the Kasner parameter is $\ue_1 = \ue_0 -2$, the
(inverse) width is $c_1$ and the time offset is ${-t_1}$ (where $t_1 \gg
t_0$). According to~\eqref{czexpr} and~\eqref{c0} we have
\begin{equation}
c_1 = -\kappa^{-1} e^t \,e^{-(t-t_1)} = c_0 e^{t-t_0} e^{-(t-t_1)} = c_0 e^{t_1 - t_0} \:,
\end{equation}
i.e., the (inverse) coordinate width increases by a factor of $e^{t_1 -
t_0}$. This establishes~\eqref{spikeevol}.

\end{appendix}


\begin{thebibliography}{99}



\bibitem{lk63} E.M.~Lifshitz and I.M.~Khalatnikov.
\newblock Investigations in relativistic cosmology.
\newblock Adv.~Phys.~{\bf 12}, 185 (1963).

\bibitem{bkl70} V.A.~Belinski\v{\i}, I.M.~Khalatnikov, and
    E.M.~Lifshitz.
\newblock Oscillatory approach to a singular point
in the relativistic cosmology.
\newblock Adv.~Phys.~{\bf 19}, 525 (1970).

\bibitem{bkl82} V.A.~Belinski\v{\i}, I.M.~Khalatnikov, and
    E.M.~Lifshitz.
\newblock A general solution of the Einstein equations with a time singularity.
\newblock Adv.~Phys.~{\bf 31}, 639 (1982).

\bibitem{khaetal85} I.M.~Khalatnikov, E.M.~Lifshitz, K.M.~Khanin, L.N.~Shur,
    and Ya~G.~Sinai.
\newblock On the stochasticity in relativistic cosmology.
\newblock {\it J.\ Stat.\ Phys.} {\bf 38}, 97 (1985). 

\bibitem{bar82} J.D. Barrow.
\newblock Chaotic behaviour in general relativity.
\newblock {\it Phys.\ Rep.} {\bf 85}, 1 (1982).

\bibitem{chebar83} D.F. Chernoff and J.D. Barrow.
\newblock Chaos in the Mixmaster Universe.
\newblock {\it Phys.\ Rev.\ Lett.} {\bf 50}, 134 (1983). 

\bibitem{waiell97} J. Wainwright and G.F.R. Ellis.
\newblock {\em Dynamical systems in cosmology}.
\newblock (Cambridge University Press, Cambridge, 1997).

\bibitem{col03} A.A. Coley.
\newblock {\em Dynamical systems and cosmology}.
\newblock (Kluwer Academic Publishers 2003).

\bibitem{aizetal97} Y. Aizawa, N. Koguro, and I. Antoniou
\newblock Chaos and Singularities in the Mixmaster Universe
\newblock {\it Progress of Theoretical Physics} {\bf 98}, 1225
(1997)

\bibitem{ber02} B.K. Berger.
\newblock Numerical Approaches to Spacetime Singularities
\newblock {\em Living Reviews in Relativity.}~{\bf 5} 6 (2002).

\bibitem{elshen87} Y. Elskens and M. Henneaux.
\newblock Ergodic theory of the mixmaster model in higher space-time dimensions.
\newblock {\it Nuc.\ Phys.\ B} {\bf 290} 111 (1987).

\bibitem{cheetal05} C. Cherubini, D. Bine, M. Bruni, and Z.
    Perjes.
\newblock The speciality index as invariant indicator in the BKL mixmaster dynamics
\newblock {\it Class.\ Quantum Grav.} {\bf 22} 1763 (2005).

\bibitem{hobetal94} D. Hobill, A. Burd, and A. Coley.
\newblock {\em Deterministic Chaos in General Relativity}.
\newblock (Plenum Press, New York, 1994).

\bibitem{corlev97} N.J. Cornish and J.J. Levin.
\newblock The mixmaster universe: A chaotic Farey tale.
\newblock {\it Phys.\ Rev.\ D} {\bf 55} 7489 (1997).

\bibitem{corlev97b} N.J. Cornish and J.J. Levin.
\newblock The mixmaster universe is chaotic.
\newblock {\it Phys.\ Rev.\ Lett.} {\bf 78} 998 (1997).

\bibitem{rugh90} S.E.~Rugh.
\newblock Chaotic Behavior and Oscillating Three-volumes in a Space Time Metric
in General Relativity.
\newblock Cand. Scient. Thesis, Niels Bohr Institute, K\o benhavn (1990).

\bibitem{motlet01} A.E. Motter and P.S. Letelier.
\newblock Mixmaster chaos.
\newblock {\it Phys.\ Lett. A} {\bf 285} 127 (2001). 

\bibitem{benmon04} R.~Benini and G.~Montani.
\newblock  Frame independence of the inhomogeneous mixmaster chaos via
Misner-Chitr\'e-like variables.
\newblock {\em Phys. Rev. D.}~{\bf 70} 103527-1 (2004). 

\bibitem{monetal08} G.~Montani, M.V. Battisti, R.~Benini and
    G.~Imponente.
\newblock Classical and Quantum Features of the Mixmaster
Singularity.
\newblock {\em Int. J. Mod. Phys.} {\bf A23} 2353 (2008).

\bibitem{damlec11a} T.~Damour and O.M.~Lecian.
\newblock Statistical Properties of Cosmological Billiards.
\newblock {\em Phys. Rev. D.}~{\bf 83} 044038 (2011).

\bibitem{damlec11b} T.~Damour and O.M.~Lecian.
\newblock About the Statistical Properties of Cosmological
Billiards.
\newblock arXiv:1103.0179,
Proceedings of The second Galileo-XuGuangqi Meeting,
11-16/07/2010, Ventimiglia, Italy (2011).

\bibitem{bel92} V.A.~Belinski\v{\i}.
\newblock Turbulence of the gravitational field near a cosmological
singularity.
\newblock Pis'ma v Zhurnal Eksperimental'noi i Teoreticheskoi Fiziki
{\bf 56} 437 (1992).

\bibitem{gow71} R.H.~Gowdy.
\newblock Gravitational Waves in Closed Universes.
\newblock {\em Phys. Rev. Lett.} {\bf 27} 826 (1971).

\bibitem{gow74} R.H.~Gowdy.
\newblock Vacuum spacetimes with two-parameter spacelike isometry groups
and compact invariant hypersurfaces: Topologies and boundary
conditions.
\newblock {\em Ann.\ Phys.} {\bf 83} 203 (1974).

\bibitem{bermon93} B.K.~Berger and V.~Moncrief.
\newblock Numerical investigation of cosmological
singularities.
\newblock {\em Phys.\ Rev.\ D} {\bf 48} 4676 (1993). 

\bibitem{grumon93} B.~Grubi{\v s}i\'c and
    V.~Moncrief.
\newblock Asymptotic behavior of the $T^3\times R$ Gowdy space-times
\newblock {\em Phys. Rev. D.}~{\bf 47} 2371�2382 (1993).

\bibitem{bermon98} B.K.~Berger and V.~Moncrief.
\newblock Evidence for an oscillatory singularity in generic
$U(1)$ symmetric cosmologies on $T^3\times R$.
\newblock {\em Phys. Rev. D.}~{\bf 58} 064023 (1998).

\bibitem{beretal98} B.K.~Berger, D.~Garfinkle, J.~Isenberg,
    V.~Moncrief, and M.~Weaver.
\newblock The Singularity in Generic Gravitational Collapse Is Spacelike,
Local, and Oscillatory.
\newblock {\em Mod. Phys. Lett.}~{\bf A13} 1565-1574 (1998).


\bibitem{bergar98} B.K.~Berger and D.~Garfinkle.
\newblock Phenomenology of the Gowdy universe on $T^3�R$
\newblock {\em Phys.\ Rev.\ D} {\bf 57} 4767 (1998).

\bibitem{beretal01} B.~K.~Berger, J.~Isenberg, and M.~Weaver.
\newblock Oscillatory approach to the singularity in vacuum spacetimes with $T^2$ isometry.
\newblock {\em Phys. Rev. D} {\bf 64} 084006 (2001).


\bibitem{her99} S.D.~Hern.
\newblock Numerical Relativity and Inhomogeneous Cosmologies.
\newblock Ph.~D. thesis, University of Cambridge (1999).
gr-qc/0004036

\bibitem{herste98} S.D.~Hern and J.M.~Stewart.
\newblock The Gowdy $T^3$ cosmologies revisited.
\newblock {\em Class.\ Quantum\ Grav.} {\bf 15} 1581 (1998).

\bibitem{renwea01} A.D.~Rendall and M.~Weaver.
\newblock Manufacture of Gowdy spacetimes with spikes.
\newblock {\em Class. Quantum Grav.} {\bf 18} 2959-2975 (2001).

\bibitem{kicren98} S.~Kichenassamy and A.D.~Rendall.
\newblock Analytic description of singularities in Gowdy
spacetimes.
\newblock {\em Class. Quantum Grav.} {\bf 15} 1339-1355 (1998).

\bibitem{ren00} A.D.~Rendall.
\newblock Fuchsian analysis of singularities in Gowdy spacetimes beyond
analyticity.
\newblock {\em Class. Quantum Grav.} {\bf 17} 3305-3316 (2000).


\bibitem{garwea03} D.~Garfinkle and M.~Weaver.
\newblock High velocity spikes in Gowdy spacetimes.
\newblock {\em Phys. Rev. D.}~{\bf 67} 124009 (2003).

\bibitem{rin10} H.~Ringstr\"om.
\newblock Cosmic Censorship for Gowdy Spacetimes.
\newblock Living Reviews in Relativity {\bf 13.2}.
http://www.livingreviews.org/lrr-2010-2 (2010).

\bibitem{waihsu89} J. Wainwright and L. Hsu.
\newblock A dynamical systems approach to Bianchi cosmologies:
orthogonal models of class A.
\newblock {\it Class.\ Quantum Grav.} {\bf 6} 1409 (1989). 

\bibitem{rin00} H.~Ringstr\"om.
\newblock Curvature blow up in Bianchi VIII and IX vacuum
spacetimes.
\newblock Class.\ Quantum\ Grav.\ {\bf 17} 713 (2000).

\bibitem{rin01} H.~Ringstr\"om.
\newblock The Bianchi IX attractor.
\newblock Annales\ Henri Poincar\'e {\bf 2} 405 (2001).

\bibitem{heiugg09a} J.M.~Heinzle and C.~Uggla.
\newblock A new proof of the Bianchi type IX attractor theorem.
\newblock Class.\ Quantum\ Grav.\ {\bf 26} 075015 (2009).

\bibitem{heiugg09b} J.M.~Heinzle and C.~Uggla.
\newblock Mixmaster: Fact and Belief.
\newblock Class.\ Quantum\ Grav.\ {\bf 26} 075016 (2009).

\bibitem{reitru10} M.~Reiterer and E.~Trubowitz.
\newblock The BKL Conjectures for Spatially Homogeneous
Spacetimes.
\newblock arXiv:1005.4908v2 (2010).

\bibitem{beg10} F.~B\'eguin.
\newblock Aperiodic oscillatory asymptotic behavior for some Bianchi
spacetimes.
\newblock Class.\ Quantum\ Grav.\ {\bf 27} 185005 (2010).

\bibitem{lieetal10} S.~Liebscher, J.~H\"arterich, K.~Webster,
and M.~Georgi.
\newblock Ancient Dynamics in Bianchi Models: Approach to Periodic
Cycles.
\newblock arXiv:1004.1989 (2010).

\bibitem{uggetal03} C.~Uggla,, H.~van Elst, J.~Wainwright and G.F.R.~Ellis.
\newblock The past attractor in inhomogeneous cosmology.
\newblock {\em Phys.\ Rev.\ D} {\bf 68}~:~103502 (2003).

\bibitem{rohugg05} N.~R\"ohr and C.~Uggla.
\newblock  Conformal regularization of Einstein's field equations.
\newblock {\em Class. Quantum Grav.}~{\bf 22} 3775 (2005). 

\bibitem{heietal09} J.M. Heinzle, C. Uggla, and N. R\"ohr.
\newblock The cosmological billiard attractor.
\newblock {\em Adv.\ Theor.\ Math.\ Phys.} {\bf 13} 293-407 (2009).

\bibitem{dametal03} T.~Damour, M.~Henneaux,  and H.~Nicolai.
\newblock Cosmological billiards.
\newblock {\em Class. Quantum Grav.} {\bf 20} R145 (2003).

\bibitem{andetal05} L.~Andersson, H.~van Elst, W.C.~Lim and
    C.~Uggla.
\newblock  Asymptotic Silence of Generic Singularities.
\newblock {\em Phys. Rev. Lett.} {\bf 94} 051101 (2005).

\bibitem{lim04} W.C.~Lim.
\newblock The Dynamics of Inhomogeneous Cosmologies.
\newblock Ph.~D. thesis, University of Waterloo (2004); arXiv:gr-qc/0410126.

\bibitem{gar04} D.~Garfinkle.
\newblock Numerical Simulations of Generic Singularities
\newblock {\em Phys. Rev. Lett.} {\bf 93} 161101 (2004).

\bibitem{lim08} W.C.~Lim.
\newblock New explicit spike solution -- non-local component of the generalized
Mixmaster attractor.
\newblock {\em Class. Quantum Grav.} {\bf 25} 045014 (2008).

\bibitem{limetal09} W.C.~Lim, L.~Andersson, D.~Garfinkle
    and F.~Pretorius.
\newblock Spikes in the Mixmaster regime of $G_2$ cosmologies.
\newblock {\em Phys. Rev. D} {\bf 79} 123526 (2009).

\bibitem{collim12} A.~Coley and W.C.~Lim.
\newblock Generating matter inhomogeneities in general relativity.
\newblock {\em Phys. Rev. Lett.} {\bf 108} 191101 (2012).

\bibitem{wai81} J.~Wainwright.
\newblock Exact spatially inhomogeneous cosmologies.
\newblock {\em J.\ Phys.\ A: Math.\ Gen.} {\bf 14} 1131-1147 (1981).

\bibitem{beretal97} B.K.~Berger, P.T.~Chru\'sciel, J.~Isenberg, and V.~Moncrief.
\newblock Global Foliations of Vacuum Spacetimes with $T^2$ Isometry.
\newblock {\em Ann.\ Phys.} {\bf 260} 117-148 (1997).

\bibitem{andetal04a} H.~Andr\'easson, A.D.~Rendall, and M.~Weaver.
\newblock Existence of CMC and constant areal time foliations in
$T^2$ symmetric spacetimes with Vlasov matter.
\newblock {\em Commun.\ PDE} {\bf 29}, 237(2004).

\bibitem{ren08} A.D.~Rendall.
\newblock Partial Differential Equation in General Relativity.
\newblock (Oxford University Press, Oxford, 2008).

\bibitem{isewea03} J.~Isenberg and M.~Weaver.
\newblock On the area of the symmetry orbits in $T^2$ symmetric spacetimes.
\newblock {\em Class.\ Quantum Grav.} {\bf 20} 3783 (2003).

\bibitem{elsetal02} H.~van~Elst, C.~Uggla, and J.~Wainwright.
\newblock Dynamical systems approach to G2 cosmology.
\newblock {\em Class.\ Quantum Grav.} {\bf 19} 51 (2002). 

\bibitem{car73} B.~Carter.
\newblock Black hole equilibrium states.
\newblock In {\em Black holes}, eds. C.~Dewitt and B..S.~DeWitt. Gordon and Breach (1973).

\bibitem{wai79} J.~Wainwright.
\newblock A classification scheme for non-rotating inhomogeneous
cosmologies.
\newblock {\em J. Phys. A\,\ Math. Gen.} {\bf 12} 2015 (1979). 

\bibitem{limetal06} W.C.~Lim, C~.Uggla and J.~Wainwright.
\newblock  Asymptotic Silence-breaking Singularities.
\newblock {\em Class.\ Quantum Grav.} {\bf 23} 2607 (2006). 

\bibitem{sanugg10} P.~Sandin and C.~Uggla.
\newblock Perfect fluids and generic spacelike singularities.
\newblock {\em Class.\ Quantum Grav.}\ {\bf 27} 025013 (2010).

\bibitem{heirin09} J.M.~Heinzle and H.~Ringstr\"om.
\newblock Future asymptotics of vacuum Bianchi type VI0 solutions.
\newblock {\em Class. Quantum Grav.}~{\bf 22} 145001 (19pp) (2005).

\bibitem{rin06} H.~Ringstr\"om.
\newblock Existence of an asymptotic velocity and implications for the asymptotic
behaviour in the direction of the singularity in $T^3$-Gowdy.
\newblock Commun.\ Pure Appl.\ Math.\ \textbf{59} 977-1041 (2006).

\bibitem{ger72} R.~Geroch.
\newblock A Method for Generating New Solutions of Einstein's Equation II.
\newblock {\em J.\ Math.\ Phys.} {\bf 13} 394-404 (1972).

\bibitem{heiuggtocome1} J.M.~Heinzle and C.~Uggla.
\newblock Generic spacelike singularities: Attractors, chains, and statistics.
\newblock Preprint (2012). 

\bibitem{chr90} P.T.~Chru\'sciel.
\newblock On Space-Times with $U(1)\times U(1)$ Symmetric Compact Cauchy Surfaces.
\newblock {\em Ann.\ Phys.} {\bf 202} 100-150 (1990).

\bibitem{andetal04} L.~Andersson, H.~van~Elst, and C.~Uggla.
\newblock Gowdy phenomenology in scale-invariant variables.
\newblock Class.\ Quant.\ Grav.\ {\bf 21} S29 (2004). 

\bibitem{ren12} A.D.~Rendall.
\newblock Dynamics of solutions of the Einstein equations with twisted
Gowdy symmetry.
\newblock {\em J.\ Geom.\ Phys.} {\bf 62} 569-577 (2012).

\bibitem{earetal72} D.~Eardley, E.~Liang, and R.~Sachs.
\newblock Velocity-Dominated Singularities in Irrotational Dust
Cosmologies.
\newblock {\em J. Math. Phys.} {\bf 13} 99 (1972). 

\bibitem{iseetal90} J.~Isenberg, M.~Jackson, and V.~Moncrief.
\newblock Evolution of the Bel-Robinson energy in Gowdy $T^3\times R$
space-times.
\newblock {\em J. Math. Phys.} {\bf 31} 517 (1990). 

\bibitem{chachr04} M.~Chae and P.T.~Chru\'sciel.
\newblock On the dynamics of Gowdy space time.
\newblock {\em Commun.\ Pure Appl.\ Math.} \textbf{57} 1015-1074 (2004).

\bibitem{rin04a} H.~Ringstr\"om.
\newblock Asymptotic expansions close to the singularity in Gowdy spacetimes.
\newblock Class.\ Quantum Grav.\ \textbf{21} S305-S322 (2004).

\bibitem{rin04b} H.~Ringstr\"om.
\newblock On Gowdy vacuum spacetimes.
\newblock Math.\ Proc.\ Camb.\ Phil.\ Soc.\ \textbf{136} 485-512 (2004).








\end{thebibliography}
\end{document}